\documentclass[aps,prd,superscriptaddress,nofootinbib,tighten,preprint]{revtex4}
\pdfoutput=1
\usepackage[utf8]{inputenc}
\usepackage[english]{babel}
\usepackage{amsmath}
\usepackage{subfigure}
\usepackage{amsfonts}
\usepackage{amssymb}
\usepackage{color}
\usepackage{cancel}
\usepackage{graphicx}
 
\usepackage{amsmath,amsfonts,amssymb}
\usepackage[hidelinks]{hyperref}
\usepackage{slashed}
\usepackage{graphicx,epsfig}
\usepackage{caption}
\usepackage{xcolor}
\usepackage{slashed,cancel,calrsfs}
\usepackage[12hr]{datetime} 
\hypersetup{
	colorlinks,
	linkcolor={blue!50!black},
	citecolor={blue!50!black},
	urlcolor={blue!80!black}
}

\newcommand{\lag}{\mathcal L}

\begin{document}
\preprint{HRI-RECAPP-2022-009}
\title{\textbf{ Distinguishing nonstandard scalar and fermionic charged particles at future $e^+e^-$ collider}}
\author{Anjan~Kumar~Barik}
\email{anjanbarik@hri.res.in}
\author{Rafiqul~Rahaman}
\email{rafiqulrahaman@hri.res.in}
\author{Santosh~Kumar~Rai}
\email{skrai@hri.res.in}
\affiliation{Regional Centre for Accelerator-based Particle Physics, Harish-Chandra Research Institute,\\ A CI of Homi Bhabha National Institute, 
Chhatnag Road, Jhunsi, Prayagraj 211019, India}

\begin{abstract}
We investigate the possibility to identify the intrinsic spin of exotic charged particles at the future $e^+e^-$ collider in 
$l^\pm+2j+\cancel{E}_T$ final state.  We choose  the Inert Doublet Model (IDM) and 
Minimal supersymmetric Standard Model (MSSM), as examples for the new physics models  
with scalar and fermionic exotic charged particles, respectively. The signal arises when these exotics charged particles are pair 
produced and then decay to a $W^\pm$ boson and the lightest neutral stable particle in the new physics model.  
We choose four benchmarks for the mass parameters which give  significant deviation from the dominant Standard Model (SM) 
$W^+W^-$ background. We find that an asymmetry in the cosine of the scattering angle 
($\cos\theta$) of one of the $W$ bosons reconstructed from $jj$ pair as well as the charged lepton have the potential to 
identify the MSSM and  IDM signal over SM with longitudinally polarized initial beams. 
A more robust distinction is seen in the shape of the azimuthal angle distribution of the $W$ boson and charged lepton, 
which can identify and distinguish the IDM signal from MSSM further if the initial beams are transversely polarized. 
\end{abstract}
\maketitle

\section{Introduction}
The Standard Model (SM) of particle physics has been a great success in explaining most of the phenomena in Nature. 
Despite its huge success, phenomena such as the existence of non-zero neutrino mass and their oscillation, the presence 
of dark matter (DM) in the universe, the obvious matter-antimatter asymmetry vis-a-vis Baryogenesis, stability of the 
electroweak scale or the gauge hierarchy problem, etc., requires one to think of physics beyond the SM (BSM). A plethora 
of candidate BSM models exist in the literature to address such non-standard phenomena. These BSM models are being 
probed at the current Large Hadron Collider (LHC), and strategies are being set up 
to probe them with more precision at future colliders such as High Luminosity LHC (HL-LHC) ~\cite{Apollinari:2015wtw}, 
High Energy LHC (HE-LHC)~\cite{Azzi:2019yne}, International Linear Collider (ILC)~\cite{Djouadi:2007ik,Baer:2013cma,Behnke:2013xla,Bambade:2019fyw}, Large Hadron electron Collider (LHeC)~\cite{LHeC:2020van}, Future Circular Collider (FCC)~\cite{FCC:2018evy,FCC:2018vvp}, etc.  
A very obvious and well known aspect of having so many different BSM theories is the so called inverse problem, 
where different models lead to overlapping outputs in the signal space, and it becomes challenging to map it to any given BSM 
scenario. Therefore the chances of observing that rare event of new physics, which has proven to be so elusive at LHC 
may have too many BSM candidates to claim as their own, since similar collider signatures can arise from different BSM 
models. Not much attention has been paid to identify the type of BSM models by looking at collider 
signatures~\cite{Asano:2011aj,Ginzburg:2014ora,Belyaev:2021ngh}. We try to fill up a gap in this direction by showing 
through this work how to distinguish two types of BSM models by looking at similar signatures at colliders based on the 
spins of exotic particles. 
Typically one expects to be able to identify spin~\cite{Boudjema:2009fz,Asano:2011aj,Christensen:2013sea,Belyaev:2016pxe} 
of a mediator by looking at the angular distribution in a $2 \to 2$ scattering process where the final state particles are SM fields. 
We consider a more common configuration which appears in models that have a DM candidate that would escape detection.
The lightest neutral stable particles (LNSP) in these models can play a good dark matter candidate. These models also 
predict charged particles that decay to LNSP. These BSM charged particles together with the LNSP can have different spins ($0$, $1/2$, $1$) in different models. Understanding the spin nature of dark matter (LNSP) experimentally is of particular importance not 
only for particle physics but also for astrophysics and cosmology. 

In this article, we investigate the possibility of identifying the spin of  BSM charged particles or the candidate dark matter through 
the collider signature of pair production of BSM charged particles followed by their decay to LNSP and $W^\pm$ boson. It 
isn't easy to do such precision measurements at the LHC as it is overwhelmed by the huge QCD background.
A future $e^+e^-$ collider or ILC, on the other hand, offers a great possibility in this direction for having a clean signature with 
a very low background and extra handles such as beam polarization (longitudinal and transverse)~\cite{MoortgatPick:2005cw}. We choose two well 
motivated models as examples, the Inert Doublet Model (IDM)~\cite{Deshpande:1977rw,Majumdar:2006nt}  and the 
Minimal supersymmetric SM (MSSM) ~\cite{Haber:1993wf,Wess:1992cp}, having potential dark matter candidates of 
type scalar (spin-$0$) and fermion (spin-$1/2$), respectively to study the collider signature of the process
\begin{equation}\label{eq:signal-proc}
e^+e^-\to C^+ \, C^-\to C^0 \, C^0 \,\, W^+ \, W^-.
\end{equation}
In IDM, the   $C^+$ and $C^0$ are $Z_2$\,-\,odd charged Higgs ($H^\pm$) and neutral Higgs ($H^0$), while for MSSM, they are the chargino ($\widetilde{\chi}^\pm$) and neutralino ($\widetilde{\chi}^0$) respectively.  Note that the charged and neutral scalars (fermions) can appear in several BSM set-ups involving 
$Z_2$ odd parity, and therefore this analysis will be applicable in all such scenarios.  We use the potential of beam 
polarization (both longitudinal and transverse) of an $e^+e^-$ collider
to discriminate between the two models by looking at various angular distributions in the $l^\pm jj +\cancel{E}_T$ final states. 

The rest of the article is organized as follows. In the next section (section~\ref{sec:models}), we briefly review the models IDM and MSSM, along with benchmark points for the new physics parameters such as masses and couplings. In section~\ref{sec:signal-background} we discuss the signal and corresponding SM background, followed by the analysis setup. In section~\ref{sec:Lpol-analysis} we study various kinematic and angular distributions with longitudinal beam polarization and perform an analysis 
with a simple cut and count on the variables. We then try to determine the spin of the exotic charged particles and their partner with the help of transverse beam polarization in section~\ref{sec:Tpol-analysis}. Finally, we conclude in section~\ref{sec:conclusion}. 

\section{Representative model}\label{sec:models}
We choose two well-motivated models, IDM and MSSM, as examples of having exotic charged particle and their neutral partner (potential dark matter) of scalar type and fermionic type, respectively. These models are briefly described below to self contain this article.
\subsection{The Inert Doublet Model}
In the Inert Doublet Model (IDM)~\cite{Barbieri:2006dq},  the scalar sector of the Standard Model (SM) is modified with 
one additional scalar  doublet $\Phi$, which is odd ($\Phi\to-\Phi$) under a new discrete $\mathbb{Z}_2$ symmetry (parity). The SM particles together with the SM Higgs doublet ($H$) are even under this $\mathbb{Z}_2$ symmetry.  
The two scalar doublets which transform under $SU(2)_L$ can be written as, 
\begin{equation}
H = \left( \begin{array}{c} G^+ \\ \frac{1}{\sqrt{2}}\left(v + h + i G^0\right) \end{array} \right),
\
\Phi = \left( \begin{array}{c} H^+\\ \frac{1}{\sqrt{2}}\left(H^0 + i A^0\right) \end{array} \right),
\end{equation}
where $v = \sqrt{2}~\langle 0 | H | 0 \rangle \approx 246$ GeV is the vacuum expectation value of the neutral component of $H$. 
The  $h$ state corresponds to the physical SM-like Higgs boson, whereas $G^0$ and $G^{\pm}$ are the Goldstone bosons. 
The ``inert'' sector consists of a neutral CP-even scalar $H^0$, a pseudo-scalar $A^0$, and a pair of charged scalars 
$H^{\pm}$. The neutral inert Higgs ($H^0$) and its charged partner ($H^\pm$) play the role of dark matter and the new exotic charged particle, respectively. 

The scalar potential of the model is given by,
\begin{align}
V & = \mu_1^2 |H|^2  + \mu_2^2|\Phi|^2 + \lambda_1 |H|^4+ \lambda_2 |\Phi|^4 
+ \lambda_3 |H|^2| \Phi|^2 + \lambda_4 |H^\dagger\Phi|^2 + \frac{\lambda_5}{2} \Bigl[ (H^\dagger\Phi)^2 + \mathrm{h.c.} \Bigr].
\label{Eq:TreePotential}
\end{align}
The masses and interactions of the scalar sector are governed by the scalar-potential parameters 
\begin{equation}
\left\{ \lambda_1, \lambda_2, \lambda_3, \lambda_4, \lambda_5, \mu_2\right\},
\end{equation} 
where $\mu_1^2$ is eliminated by $M_{h}^2 = -2\mu_1^2 = 2 \lambda_1 v^2$ which is obtained by minimizing the scalar potential after EWSB.
The IDM parameter space can be  expressed in terms of  physically more intuitive set
\begin{equation}
\left\{ M_{h}, \ M_{H^0}, \ M_{A^0}, \ M_{H^{\pm}}, \ \lambda_L, \ \lambda_2 \right\},
\label{eq:masses}
\end{equation} 
where the Higgs and inert scalar masses are given by
\begin{align}
M_{h}^2 &= \mu_1^2 + 3 \lambda_1 v^2, \\ 	
M_{H^0}^2 &= \mu_2^2 + \lambda_L v^2, \label{Eq:mH0tree} \\
M_{A^0}^2 &= \mu_2^2 + \lambda_S v^2, \\
M_{H^{\pm}}^2 &= \mu_2^2 + \frac{1}{2} \lambda_3 v^2, 
\end{align}
and the couplings $\lambda_{L},\lambda_{S}$ are defined as
\begin{eqnarray}
\lambda_{L} &=& \frac{1}{2} \left( \lambda_3 + \lambda_4 + \lambda_5 \right)\nonumber \\ 
\lambda_{S} &=& \frac{1}{2} \left( \lambda_3 + \lambda_4 - \lambda_5 \right).
\end{eqnarray}

The interaction Lagrangian in our signal  comprising the production and decay vertices of $H^\pm$ originating from 
gauge interactions can be written as,
\begin{eqnarray}\label{eq:intlag-idm}
{\mathcal L}
&=&
i \left[ g_Z (1/2 - s_W^2) Z^\mu + e A^\mu \right]
\left[
\left(\partial_\mu H^+_S \right) H^-_S
-
\left(\partial_\mu H^-_S \right) H^+_S
\right] 
\nonumber \\
&&
+(g/2)
\left[
-\left(\partial^\mu H^+_S \right) H^0_S W^-_\mu
+\left(\partial^\mu H^0_S \right) H^+_S W^-_\mu
+ h.c.
\right].
\end{eqnarray}
Here, $e=\sqrt{4\pi\alpha}$ with  $\alpha$ being fine structure constant, $g$ as the $SU(2)_L$ gauge couplings,  $g_z=e/(s_Wc_W)$, $s_W=\sin\theta_W$, and $c_W=\cos\theta_W$ with $\theta_W$ being the Weinberg  angle. 

\subsection{Minimal supersymmetric SM}
We briefly discuss the basic setup of the model and the relevant spectrum used in our analysis.
The Minimal Supersymmetric SM (MSSM)~\cite{Drees:2004jm} is the supersymmetric extension of SM 
where for every SM fermion (boson), there is a boson (fermion) superpartner. The MSSM has two Higgs doublet superfields 
with opposite hypercharge (needed to cancel the resulting gauge anomaly). 
Supersymmetry (SUSY) cannot be an exact symmetry as it would lead to similar  mass for the superpartners of the SM particles.  
SUSY  is softly broken, and in MSSM through explicit mass terms for the superpartners of the SM particles. 
The Lagrangian in MSSM has a discrete global symmetry called R-Parity ($R_{P}$), defined as $(-1)^{3B-L+2S}$ whereas 
all SM particles are even under $R_P$ while their superpartners have $R_P=-1$. This makes  the lightest SUSY particle (LSP) 
stable and is considered as a dark matter (DM) candidate. For our analysis, we consider the LSP as a composition of  
the higgsino and gaugino states of MSSM. The Lagrangian containing only higgsinos and gauginos  is given by,
\begin{eqnarray}
\lag & \supset & i \tilde{H}_{1i} \sigma^{\mu} {\Delta_{\mu}}_{ij} \bar{\tilde{H}}_{1j} + i \tilde{H}_{2i} \sigma^{\mu} {\Delta_{\mu}}_{ij} \bar{\tilde{H}}_{2j} + \mu \epsilon_{ab} \tilde{H}_{1a} \tilde{H}_{2b} + i \tilde{B} \sigma^{\mu} \partial_{\mu} \bar{\tilde{B}}+i \lambda_{i} \sigma^{\mu} {\Delta_{\mu}}_{ij} \bar{\lambda}_{j} \nonumber \\
&-& \frac{1}{2} M_{1} \tilde{B} \tilde{B} - \frac{1}{2} M_{2} \lambda_{i} \lambda_{i} 
-  \sqrt{2}  g_{2}  \tilde{H}_{1i}\lambda_{a} \frac{\sigma^a_{ij}}{2} {H}_{1j}  - \sqrt{2}  g_{2}  \tilde{H}_{2i}\lambda_{a} \frac{\sigma^a_{ij}}{2} {H}_{2j}\nonumber \\
&-&   \sqrt{2}  g_{1}\frac{\hat{Y}_{i}}{2} \tilde{B} \tilde{H}_{1i} {H}_{1i}  -  \sqrt{2}  g_{1}\frac{\hat{Y}_{i}}{2} \tilde{B}\tilde{H}_{2i} {H}_{2i} + h.c.,
\end{eqnarray}
where 
\begin{align*}
  \Delta^{\mu}_{ij} = \delta_{ij} \partial^{\mu} + i g_{1} \frac{\hat{Y}_{i}}{2}\delta_{ij}\, B_{\mu} + i g_{2} W^{\mu a}\,T^a_{ij}
\end{align*}
Here $\hat{Y}$  and $T^a$ respectively represents hypercharge and  $SU(2)$ generators in the respective representation of the field over these operator act.

and 
\begin{align*}
  \tilde{H}_{1} = \left( \begin{array}{c}  \tilde{h^{1}_{1}} \\ \tilde{h^{2}_{1}}\end{array} \right), && 
  \tilde{H}_{2}= \left( \begin{array}{c}  \tilde{h^{1}_{2}} \\ \tilde{h^{2}_{2}}\end{array} 
  \right),  && \tilde{B}, && \lambda_s  = (\lambda_1,\lambda_2,\lambda_3)
\end{align*}
are the fermionic super partner of two Higgs fields $H_1$, $H_2$ and the super partners of $U(1)$ , $SU(2)$  gauge fields, 
 respectively. 
Here  the neutral sector of SUSY fermionic partners of the SM bosons consist of neutral Weyl fermions 
($\tilde{B}$, $\lambda_{3}$, $\tilde{h}^{1}_{1}$, $\tilde{h}^{2}_{2}$) and the charge sector comprises 
of ($\tilde{h^{1}_{2}}$, $\tilde{h^{2c}_{1}}$, 
$\lambda^{+} \equiv \frac{\lambda_1 - i \lambda_2}{\sqrt{2}}$, $\lambda^{-c} \equiv \frac{\lambda^c_1 - i \lambda^c_{2}}{\sqrt{2}}$, superscript `c'   being 
the charge conjugation operator) two component Weyl fermions. 

Spontaneous symmetry breaking is realised through vacuum expectation value (VEV) for the two Higgs fields 
$H_1$ and $H_2$ with VEV $\langle \frac{v_1}{\sqrt{2}} \rangle$ and $\langle\frac{v_2}{\sqrt{2}} \rangle$  respectively.  
Here, we define Electro-Weak VEV $v = \sqrt{v^2_1+v^2_2}$ with   $\tan\beta = \frac{v_2}{v_1}$.

After electroweak symmetry breaking, the mass term for the higgsino and Electro-Weakino charged sector becomes
\begin{equation}
\lag^{c}_{Mass} = -\frac{g_{2}}{\sqrt{2}}(v_{1}\lambda^{+}\tilde{h^{2}_{1}}+v_{2}\lambda^{-}\\\tilde{h^{1}_{2}} + h.c.) - (M_2 \lambda^{+} \lambda^{-} + \mu \tilde{h^{2}_{1}} \tilde{h^{1}_{2}} + h.c.).
\end{equation}
In the basis $\psi^{+} = \left( \begin{array}{c} \lambda^{+} \\ \tilde{h^{1}_{2}}\end{array} \right)$ and 	$\psi^{-} = \left( \begin{array}{c} \lambda^{-} \\ \tilde{h^{2}_{1}}\end{array} \right)$, 
the above Lagrangian can be written as 
\begin{equation}
-\lag^{c}_{Mass} = (\psi^-)^T\begin{pmatrix}
M_2 & \sqrt{2}M_W sin\beta\\
\sqrt{2}M_W cos\beta & \mu
\end{pmatrix}\psi^+.
\end{equation} 
As the mass matrix is not symmetric, it has to be diagonalized by a bi-unitary transformation, 
\begin{equation}
M^{D}_{c} = \mathcal{U}^*\begin{pmatrix}
M_2 & \sqrt{2}M_W sin\beta\\
\sqrt{2}M_W cos\beta & \mu
\end{pmatrix} \mathcal{V}^{-1} ,
\end{equation} 
where $M^{D}_{c}$ is a diagonal matrix with real positive eigenvalues. Weyl fermion eigenstates will be 
$\chi^+_k = \mathcal{V}_{km}\psi^+_{m}$ and $\chi^-_k = \mathcal{U}_{km}\psi^-_{m}$, 
which can be written as four component chargino field. The mass for the charginos can be written as
\begin{equation}
-\lag^c_{Mass}=\tilde{M_1 }\overline{\tilde{\chi}^+_1}\tilde{\chi}^+_1 + \tilde{M_2 } \overline{\tilde{\chi}^+_2}\tilde{\chi}^+_2, 
\end{equation} 
where $\tilde{\chi}^+_1\equiv \begin{pmatrix}
\tilde{\chi}^+_1\\
\overline{\tilde{\chi}^-_1}^T
\end{pmatrix}$ and  $\tilde{\chi}^+_2\equiv \begin{pmatrix}
\tilde{\chi}^+_2\\
\overline{\tilde{\chi}^-_2}^T
\end{pmatrix}$ are chargino fields. The Lagrangian in the neutral electroweakino sector can be written as 
\begin{equation}
\lag^n_{Mass} = -\frac{g_2}{2}\lambda_{3}(v_{1}\tilde{h}^1_1 - v_{2}\tilde{h}^2_2 ) +\frac{g_1}{2}\tilde{B}(v_{1}\tilde{h}^1_1 - v_{2}\tilde{h}^2_2 ) + \mu \tilde{h}^{1}_{1} \tilde{h}^{2}_{2} -\frac{1}{2} M_2 \lambda_{3}\lambda_{3} - \frac{1}{2} M_1 \tilde{B}\tilde{B}+h.c.
\end{equation}
The mass matrix, in the basis of $\psi^0 = (\tilde{B},\lambda_{3},\tilde{h}^{1}_{1},\tilde{h}^{2}_{2})$ is given by 
\begin{equation}
\mathcal{M}^n=\begin{pmatrix}
M_1 & 0 & - M_z c_\beta s_W & M_z s_\beta s_W\\
0 & M_2 &  M_z c_\beta c_W & - M_z s_\beta c_W \\
- M_z c_\beta s_W &  M_z c_\beta c_W & 0 & -\mu \\
M_z s_\beta s_W & - M_z s_\beta c_W & -\mu & 0
\end{pmatrix}.
\end{equation}
This  matrix can be diagonalized by a unitary matrix $Z$. The physical mass eigenstates and the mass diagonalization matrix are
given by $\chi^0_l = Z_{ln}\psi^0_n$ and  $Z^*\mathcal{M}Z^{-1} = M^n_D$, respectively. In the four component notation the 
mass term in the Lagrangian for neutralinos can be written as
$\lag^n_{Mass}= - \frac{1}{2}\sum_{l} \tilde{M}^n_l \overline{\tilde{\chi}^0_l}\tilde{\chi}^0_l  $, where  $\tilde{\chi}^0_l = \begin{pmatrix}
\chi^0_1\\
{\overline{\chi}^0_l}^T
\end{pmatrix} $ are the four Majorana neutralino fields. The lowest mass eigenstate of four neutralinos will be the LSP 
and represents the DM candidate.

The masses and vertices relevant for our analysis except the gauge couplings depend only on four parameters which are 
the soft breaking gaugino mass parameters ($M_1$ and $M_2$), the higgsino mass parameter ($\mu$) and 
the ratio of the VEV of the Higgs doublets ($\tan\beta$). The lightest chargino mass eigenstate  coupling to $Z$ boson 
depends on the chargino  mixing matrix elements $\mathcal{V}_{11}$ and $\mathcal{U}_{11}$ ~\cite{Choi:1998ut}. 
For simplicity we keep  these matrix elements fixed for  all our benchmark points which can be obtained by varying the 
input parameters $M_2$, $\mu$ and $\tan\beta$ of the model.  

\subsection{Benchmark selection}
\begin{table}[h]
	\centering
	\begin{tabular}{|c|c|}\hline
		Benchmark& Masses  \\ \hline
		BP1&$M_{\pm}=160$ GeV, $M_{0}=60$ GeV \\ \hline
		BP2& $M_{\pm}=220$ GeV, $M_{0}=100$ GeV \\ \hline
		BP3& $M_{\pm}=220$ GeV, $M_{0}=120$ GeV \\ \hline
		BP4& $M_{\pm}=300$ GeV, $M_{0}=10$ GeV \\ \hline
	\end{tabular}
\caption{\label{tab:BP} Benchmark points for the MSSM and IDM. }
\end{table}
We choose four benchmark points ({\tt BP}s), in our analysis, for the masses of exotic charged particles ($C^\pm$) 
and their neutral partner ($C^0$) in both IDM and MSSM. The benchmark points are listed in Table~\ref{tab:BP}. 
The {\tt BP}s are chosen in such a way that $\Delta M=M^+-M^0$ ($M^+=M_{C^+}$, $M^0=M_{C^0}$) remain fixed 
for two {\tt BP}s with different $M^+$ and $M^0$ ({\tt BP1} and {\tt BP3}); Two {\tt BP}s have the same $M^+$ but 
with a  different $\Delta M$ ({\tt BP2} and {\tt BP3}); Three {\tt BP}s satisfy these criteria.  We choose one 
more {\tt BP} which has a different $M^+$ and $\Delta M$ compared to the other three ({\tt BP4}). 

The mass parameters are kept the same in both models in order to have similar kinematic behavior. Through all four {\tt BP}s, the 
mass difference between charged odd particle with respective DM particle is kept higher than the mass of $W$-boson, so that 
both $W$ boson and DM particle can be produced on-shell from their parent charged dark sector particle. 
In IDM, there are two coupling parameters, $\lambda_L$ and $\lambda_S$, which can affect the SM 
Higgs signal in our study. The value of  $\lambda_L$ is kept very small so that the second neutral CP-even Higgs mass ($m_{H_0}$) which is
the DM,  is  nearly equal to $ \mu_2$ (see, Eq.~(\ref{Eq:mH0tree})). The value of $\lambda_S$  is chosen such that 
$m_{A_0} - m_{H^+} = 1 $ GeV. These two choices help us in evading the bound on Higgs invisible decay as well as  electroweak  
precision observables.  For all four {\tt BP}s of MSSM we have kept the  chargino mixing matrices to be the same 
with $\mathcal{V}_{11}\equiv\cos\theta_R=0.9725$ and  $\mathcal{U}_{11}\equiv\cos\theta_L=0.9168$. 
We do not focus on whether the benchmarks satisfy  the requirements of dark matter relic density. However, we 
keep {\tt BP1} as a reference point which does satisfy  dark matter constraints \cite{Belyaev:2021ngh}. 
Apart from the dark matter constraint, there are other constraints that are considered while 
choosing the {\tt BPs}, which we discuss below.
\textcolor{black}{\subsection{Constraints}
\subsubsection{\textbf{Vacuum stability and unitarity }}
In IDM, the scalar potential must be bounded from below. The condition on various quartic couplings to satisfy this constraint are given by~\cite{Belyaev:2016lok}
 \begin{align}\label{eq:stability}
 &\lambda_1 > 0, \,\,\,\,  \lambda_2 > 0, &&  2\sqrt{\lambda_1 \lambda_2 } + \lambda_3 > 0, && 
 2\sqrt{\lambda_1 \lambda_2 } + \lambda_3 + \lambda_4 - |\lambda_5| > 0  .
 \end{align}
  We have checked that all our four {\tt BPs} for IDM satisfy the above conditions. In addition, we have also checked that the {\tt BPs} 
   satisfy unitarity bounds~\cite{Belyaev:2016lok}.
\subsubsection{\textbf{Electroweak  precision observables}}
The BSM particles affect the electroweak (EW) observables via oblique correction and these corrections are parameterized 
by three observables  $S, \, T$ and $U$ as electroweak  precision observables (EWPO). The contribution to EWPO ($S$ and $T$) 
from IDM are given by~\cite{Belyaev:2016lok}
\begin{equation}\label{eq:ewpo-S}
 S = \dfrac{1}{{72 \pi  \left(x_2^2-x_1^2\right)^3}}
\left[ x_2^6 f_a(x_2)-x_1^6 f_a(x_1)+9x_2^2x_1^2\left( x_2^2f_b(x_2)-x_1^2f_b(x_1) \right) \right],
\end{equation}
where $x_1=\frac{M_{H^0}}{M_{H^\pm}}$, $x_2=\frac{M_A}{M_{H^\pm}}$, $f_a(x)=-5+12\log (x)$, $f_b(x)=3-4\log (x)$ and
\begin{equation}\label{eq:ewpo-T}
T=\dfrac{1}{{32 \pi ^2 \alpha  v^2}} \left[f _c\left(M_{H^+}^2,M_A^2\right)-f_c\left(M_A^2,M_{H^0}^2\right)+f_c\left(M_{H^+}^2,M_{H^0}^2\right)\right].
\end{equation}
where
\begin{align*}
f_c(x,y) =
\begin{cases}
\frac{x+y}{2}-\frac{x y \log \left(\frac{x}{y}\right)}{x-y}, & {x\neq y}\\
0, & {x = y}.
\end{cases}  
\end{align*}
 The experimental values of $S$ and $T$ with $U = 0$ is given by \cite{Haller:2018nnx}
\begin{equation}
S = 0.04\pm 0.08  ,\,\, {\rm and} \,\, T =  0.08\pm 0.07 .
\end{equation}  
 In our case, {\tt BP4} with the largest  $\Delta M =M_{H^\pm} -M_{H^0}$ in IDM gives the maximum deviation in 
 EWPO with $\Delta S = -\,0.0218 $ and $ \Delta T = -\,0.0017 $. These values are well within the $ 1 \sigma$ limit 
 of experimental uncertainty.
\subsubsection{\textbf{Higgs invisible decay}}
We note that only {\tt BP1} and {\tt BP4} have $m_h>2 \, m_{DM}$. Hence the SM Higgs can decay to two DM particles for these 
benchmark points. The observed value of Higgs invisible branching fraction is less than 0.18 at 
$95 \%$ C.L. ~\cite{CMS:2022qva}.
For MSSM, our branching fraction ($BR(h\to 2\bar{\chi}^0)$) is around $2\,\% $ and $4\,\% $ for  {\tt BP1} and {\tt BP4} respectively. 
In IDM, the value of  $\lambda_L$ which enters into the interaction strength is kept fixed at $10^{-4}$ for all {\tt BP}s, which 
helps us keep the invisible branching ratio (BR) of $h \to H_0\, H_0$ below $1\,\% $ for both {\tt BP1} and {\tt BP4}.
\subsubsection{\textbf{Higgs Signal}}
The charged Higgs and chargino being at the electroweak scale, can contribute to the loop-induced $h\to\gamma\gamma$ 
	decay. This can affect the Higgs signal observations at the LHC in this channel. The observed value of the signal strength 
	is constrained by ATLAS~\cite{ATLAS:2022tnm} as
$$\mu_{\gamma\gamma} = \frac{BR_{BSM}(h\to\gamma\gamma)}{BR_{SM}(h\to\gamma\gamma)}= 1.04_{-0.09}^{+0.10} \,\, . $$
For IDM, the maximum contribution to $BR(h\to\gamma\gamma)$ comes from BP1 which has the lightest $H^+$, and the 
total Higgs signal strength in this particular channel with respect to the SM predictions i.e., 
$\mu_{\gamma\gamma}^{\rm IDM} = 0.975$. The deviation from the SM prediction goes down as the charged Higgs mass increases.
%
%
Similarly for MSSM, {\tt BP1} gives the maximum contribution	to $BR(h\to\gamma\gamma)$ with  
	$ \mu_{\gamma\gamma}^{\rm MSSM} =1.21$. We find that the Higgs signal strength for both IDM and MSSM are 
	within $1 \sigma $ and  $2 \sigma $  of the allowed signal strength, respectively.  }
\section{Signal and background}\label{sec:signal-background}
\begin{figure}[h]
	\centering
	\includegraphics[width=0.4\textwidth]{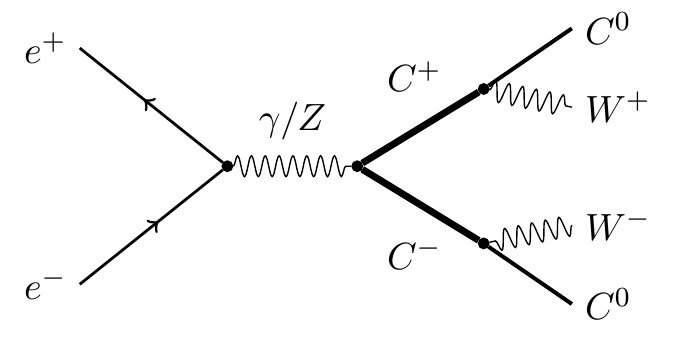}
	\caption{\label{fig:feyndia} Representative Feynman diagram for the signal process $e^+ \, e^- \rightarrow  C^+ \,  C^- \rightarrow W^+ \,  W^- \, 2C^0$.}
\end{figure}
The signal for our analysis comes from the pair production of the exotic charged particles ($C^\pm$) followed by their decay to 
their neutral partner ($C^0$) and charged gauge boson $W^\pm$. It is worth pointing out here that most models, which are 
proposed to give a DM candidate in the spectrum exhibit a similar chain of production where the final state involves a pair of 
DM particles. The visible particle multiplicity associated with the two DM final states depends on the particle being produced 
in the hard scattering process. The challenging aspect of obtaining good signal sensitivity then relies on the strength 
with which the hard scattering process takes place. Producing the electroweak strength particles at LHC leads to very 
weak sensitivities and therefore makes the spin determination very difficult. We, therefore, consider their production at ILC
in the simplest and most obvious mode of production process.
We choose the channel where one of the $W$ decays leptonically while the other one decays hadronically, i.e.,
\begin{equation}
e^+e^-\to C^+C^-,~C^\pm\to C^0 W^\pm,~W^\pm\to l^\pm \nu_l/\bar{\nu_l}, ~W^\mp\to jj,
\end{equation}
forming a final state of $l^\pm 2j+\slashed{E}_T$.
The Feynman diagram up to $W^\pm$ production is shown in Fig.~\ref{fig:feyndia}.
The signal process in IDM and MSSM, along with the SM background, are described below. 

\paragraph{\rm\bf  IDM Signal:}
In IDM, $H^+ H^- $ is produced via two $s$-channel processes with $Z$ boson and $\gamma$ as mediators from initial $e^- e^+$. 
Then $H^{\pm}$ decay to $W^{\pm} H_0$ where $H_0$ is stable while the $W$ boson decays to lepton and hadrons 
producing one lepton, two jets and missing energy final states. We have kept the pseudoscalar  $A_0$ mass higher than 
$H^{\pm}$. Since there are no other lighter particles which are $Z_2$ odd,   branching ratio (BR) of  $H^{\pm}$  
to $W^{\pm} H_0$ channel will be $100 \%$. The  two DM particles $H_0$ and neutrino act as the source of missing energy.

\paragraph{\rm\bf MSSM Signal: }
In MSSM, similar $s$-channel diagrams can produce the lightest chargino $\chi^{\pm}_1$ at ILC. Although there is a 
possibility of a $t$-channel diagram due to a sneutrino, we assume that all scalar superpartner masses are very heavy including the 
sneutrino, which is kept at a  mass of 12 TeV. Hence the contribution from the sneutrino exchange will be much suppressed 
due to the large mass of sneutrino. The charginos then decay to $W^{\pm} \chi_{0}$ and further decay of $W^{\pm}$  
as before, giving the desired final states with the neutralinos now playing the role of DM. 
For our analysis, we have taken all $R_P$ odd particles except the lightest neutralino heavier than the mass of 
$\chi^{\pm}_1$ which again gives its decay  BR to $W^{\pm}\, \chi_{0}$ as  $100 \%$. Here too the  missing energy 
gets contribution from the two DM states $\chi_0$ and a neutrino.

\paragraph{\rm\bf SM Background:}
The dominant background for our signal is the $W^+W^-$ production in SM where one $W$ boson decays 
leptonically while the other one decays hadronically. There are three major sub-processes in SM contributing to 
$W^- W^+$ production, with two being $s$-channel 
processes having  photon and $Z$ boson propagator and the other and most dominant  $t$-channel process with the light 
neutrino running in the propagator. The $W^- W^+ Z$ production with $Z\to\nu\bar{\nu}$ also produces the same final state, 
but this background is negligibly small (roughly a hundred times smaller in cross section than $W^+W^-$ production) 
due to an additional electroweak coupling and phase-space suppression.  Additionally the  small branching for
 $Z\to\nu\bar{\nu}$ decay also makes the contributions from this process weaker. 
 We can therefore safely ignore the $W^- W^+ Z$ contributions for the background in our analysis. 

We wish to study the polarization and spin of the intermediate exchanged as well as produced 
particles. We have generated the signal and background events in the package {\tt WHIZARD}~\cite{Kilian:2007gr} at 
ILC with $\sqrt{s}=1$ TeV center of mass energy. We generate the events with the initial state radiation (ISR) effect switched on. 
During the event generation, we have put a set of inclusive cuts with $P_{{T}_{j_1,j_2,l}} > 10$ GeV and the invariant mass of 
jets $M_{jj}$  to lie  between $60$ GeV to $100$ GeV. Then the simulated events are showered 
in {\tt Pythia8}~\cite{Sjostrand:2014zea} for energy smearing effects. After the hadronization of the final state jets we perform 
the fast detector simulation in {\tt Delphes-3}~\cite{deFavereau:2013fsa} with the International Linear Detector (ILD) card. 
Events are selected at the detector level with the following selection cuts,
\begin{eqnarray}\label{eq:sel-cut}
&p_{T_j}>30~\text{GeV},~p_{T_l}>30~\text{GeV},~\cancel{E}_T>30~\text{GeV},~60~\text{GeV}<M_{jj}<100~\text{GeV},&\nonumber\\
&~|\eta_j|<4.5,|\eta_l|<2.5,
~\Delta R_{j,j}>0.5~,\Delta R_{l,l}>0.5,~\Delta R_{l,j}>0.5.&
\end{eqnarray}
We name these cuts as {\tt Sel\_cut}.

\section{Signal analysis at ILC with beam polarization}
We now present our signal analysis using polarized beams in the initial states. Initial-state polarization is a useful 
diagnostic at the ILC and by  adjusting initial-state polarizations, one can select specific states 
preferentially.  In this work, we present our analysis by using the option of using both $e^-$ and $e^+$ beams longitudinally 
polarized as well as having them transversely polarized. Since we want to show how the spin of the newly produced 
particles could show up in some kinematic variables, it is useful to find out which polarization option will be best 
suited to help identify such states with better confidence. 

Unlike the LHC, where the parton collisions are over a range in center of mass energies, we know that ILC will have a 
more or less fixed center of mass energy for the $e^-e^+$ collisions. A small spread in the 
collision energy however arises from Initial State Radiation (ISR) at the ILC, which  is the most important QED correction to the 
Born cross section~\cite{Yennie:1961ad}.  Bremsstrahlung effects are an important source of ISR, and it is the radiation 
caused by the interaction between the electron and positron participating in the annihilation event at $e^+e^-$colliders. 
Thus its effects must be considered for realistic simulation to study physics signals at future linear colliders such as the ILC.  
The radiative corrections to processes with arbitrary final states need evaluation to achieve precision measurements. 
The ISR photons are generally soft with small transverse momenta, so they eventually escape detection. However, their effect is imprinted in the physics analysis through modification of the colliding beam energies 
and an effective boost along the beam axis for the final states. For a realistic analysis we, therefore, include the 
ISR effects in our study.  A somewhat subdominant correction for the center of mass energy also comes from another
phenomenon called beamstrahlung which we have neglected. 

\subsection{Analysis with longitudinal beam polarization}\label{sec:Lpol-analysis}
\begin{table}[h!]
	\begin{center}\scalebox{1.0}{
			\begin{tabular}{|c|c|c|c|c|c|}
				\cline{3-6}
				\multicolumn{2}{c|}{}&\multicolumn{2}{c|}{IDM Cross section (fb)}&\multicolumn{2}{c|}{MSSM Cross section (fb)} \\ \hline
				Benchmark      & Mass($M^{\pm}$, $M_0$) (GeV) & Un-pol  &  Lpol    & Un-pol   &   Lpol        \\ \hline 
				{\tt BP1}   &  $(160, 60)   $                & $5.911 $  &  $14.331 $ & $55.163  $ &  $ 158.900  $ \\
				{\tt BP2}   &  $(220, 100)  $                & $4.996 $  &  $12.103 $ & $54.959  $ &  $ 158.136  $ \\
				{\tt BP3}   &  $(220, 120)  $                & $5.006 $  &  $12.146 $ & $54.799  $ &  $ 157.702  $ \\
				{\tt BP4}   &  $(300, 10)   $                & $3.699 $  &  $8.966  $ & $54.445  $ &  $ 156.363  $ \\
				\hline 
		\end{tabular}}
	\end{center}
	\caption{The cross sections of signal  for the final state $e^- \, e^+ \rightarrow  l^\pm+2 j+\slashed{E}_T$ after 
		the basic generation level cuts.}
	\label{tab:BPscross}
\end{table}
We perform our collider analysis in the chosen final state with longitudinal beam polarization (LPol) of initial 
$e^{\pm}$ beams primarily because of the larger cross sections than with the unpolarized (Un-Pol) beams;
The degree of polarization for ($e^-$, $e^+$) beams are chosen to be  ($-\, 80\, \% $ ,  $+\, 60\, \% $ ). The SM $WW$ background has a cross section of about $680.954$ fb in Un-Pol, and  $1957.84$ fb in LPol with generation level cuts given in Eq.~(\ref{eq:sel-cut}). The larger cross sections are effectively due to the choice of polarization that enhances left-handed current contribution.

The cross sections of the signal are larger in LPol too, compared to Un-Pol  in all four {\tt BP}s, shown in Table~\ref{tab:BPscross}, by roughly the same factor by which  the  background cross section is larger in LPol compared to Un-Pol. Thus, signal significance improves in LPol than in Un-Pol, even without any kinematic cuts to reduce SM background. 
From Table~\ref{tab:BPscross}, it follows that IDM has $\mathcal{O}(1)$ less cross section than MSSM. \textcolor{black}{This can be understood in the following way. In the massless limit (boost $\beta\to 1$), for the photon mediated diagram, the total pair production cross section of $ H^\pm$ is four times smaller than the pair production cross section of $ \chi_1^\pm$  at an electron-positron collider. The reason for the enhanced cross section of $ \chi_1^\pm$ pair production is that there are four ways of combining helicity states of $ \chi_1^\pm$ and $ e^\pm$,   while for scalar there is only one degree of freedom for  $ H^\pm$ each. The production cross section for a pair of fermions again goes up when we consider them to be massive. 
	For the photon mediated diagram, the cross sections follow as 
	\begin{equation}
	\sigma(e^- e^+ \to H^- H^+)\propto \frac{2}{3}e^4\beta^3 \,\, \text{and}\,\, \sigma(e^- e^+ \to \chi_1^- \chi_1^+)\propto -\frac{4}{3}e^4\beta(-3+\beta^2)
	\end{equation}	
	for unpolarized initial beams (see Eqs.~(\ref{eq:idm-photon}), (\ref{eq:mssm-photon}), and (\ref{eq:dsig-dphi})).
	Similarly for the diagram with $Z$-boson propagator, the factor associated with $Z$-boson vertex $(T_3 - \sin^2\theta_W \, Q ) $ is larger for chargino than the charged Higgs. In our case, the lightest chargino is wino dominated. Following the notation of Appendix~\ref{app:tpol-analytics}, the ratio between the production cross section of the chargino pair and the charged Higgs pair is \,$4\left(a_\chi^2+v_\chi^2\right)/\left(g_2 \cos{\theta_W} -g_1 \sin{\theta_W}\right)^2 = 27.24$  for unpolarized beam in $\beta \rightarrow 1$ limit. The individual cross section from the photon and the $Z$ diagrams are of the same order and they 
	interfere destructively. Considering both the photon and the $Z$ mediated diagram, we get chargino pair production cross section nearly $8$ times larger than the charged Higgs pair production cross section for unpolarized beam in $\beta \rightarrow 1$ limit. This ratio increases further when we consider the final state particles to be massive.}
	However, it should be noted that in generic scenarios not restricted to SUSY where interactions with such exotics emerge from suppressed mixing angles,
smaller branching fractions in the decay modes or additional subprocesses contributing destructively can lead to somewhat comparable cross sections with the scalar production. Nevertheless, the large cross section for the fermion production is always a good discriminator when compared to scalar production in simple setups.
\paragraph{\rm \bf Kinematic variables}
\begin{figure}[h!]
	\centering
	\includegraphics[width=0.45\textwidth]{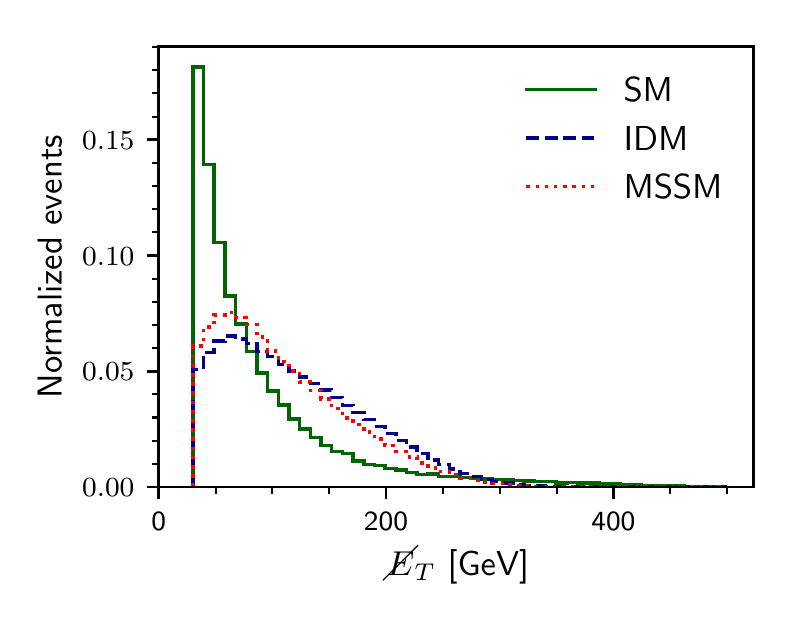}
	\includegraphics[width=0.45\textwidth]{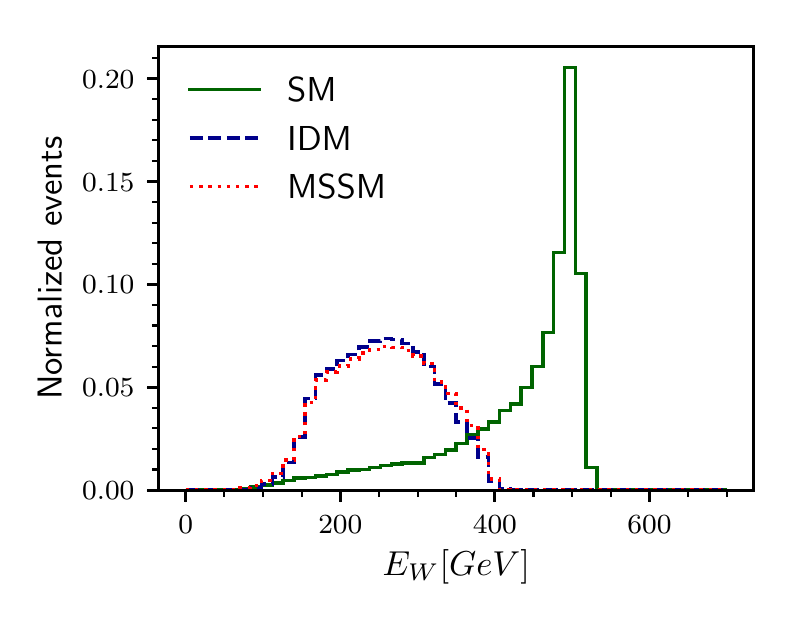}
    \includegraphics[width=0.45\textwidth]{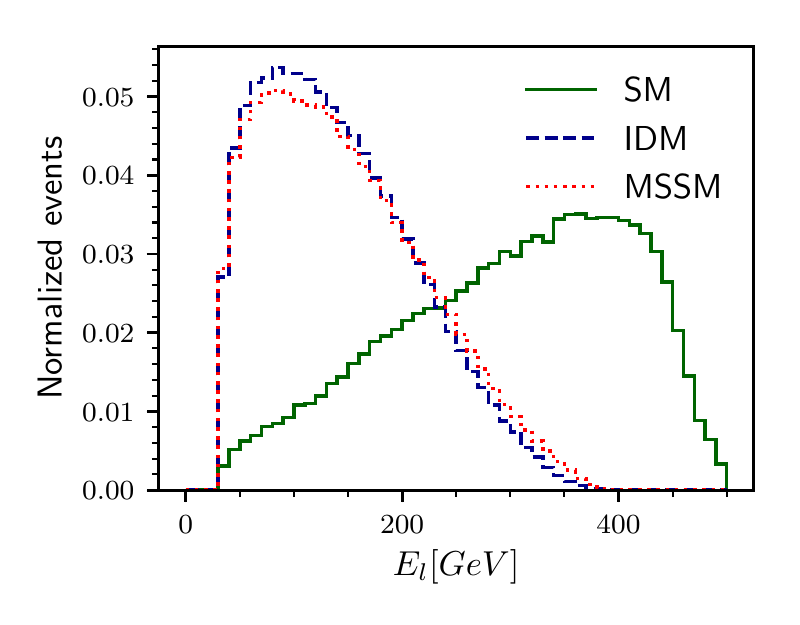}	
	\includegraphics[width=0.45\textwidth]{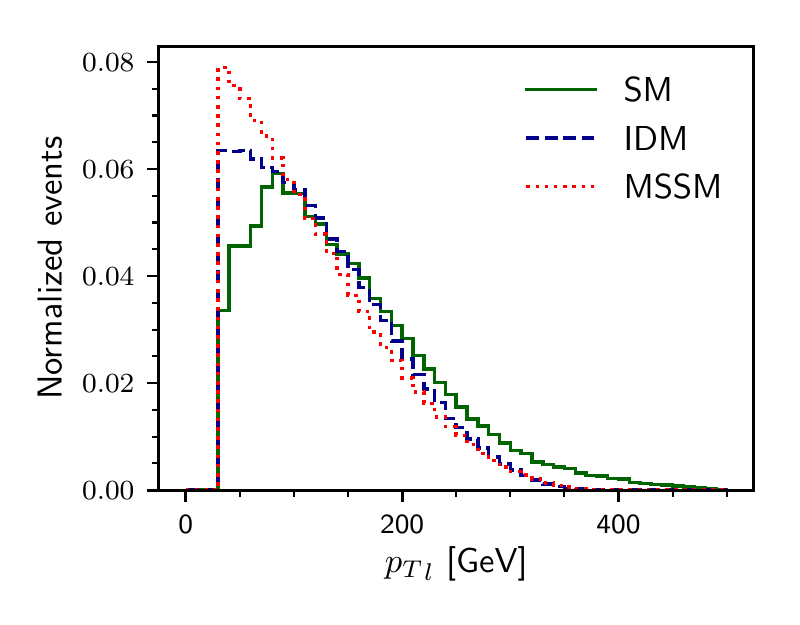}
	\includegraphics[width=0.45\textwidth]{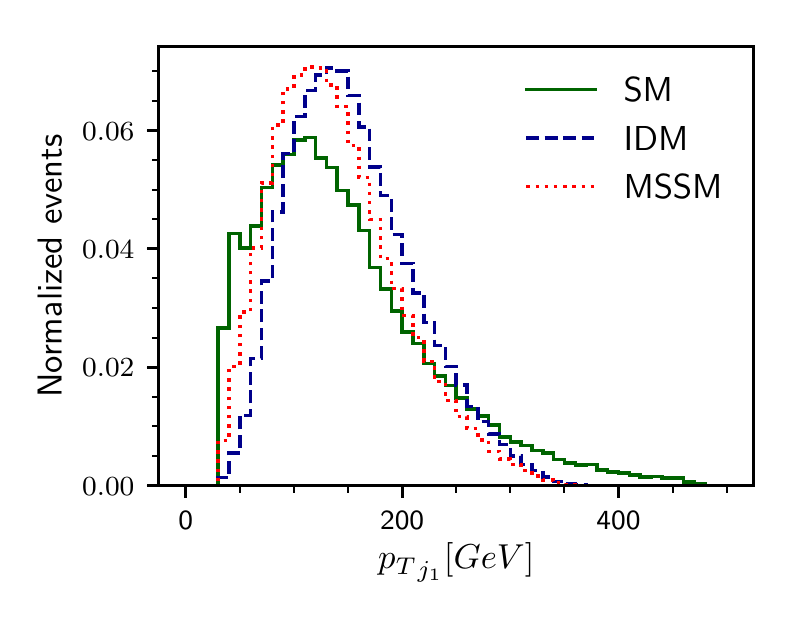}
	\includegraphics[width=0.45\textwidth]{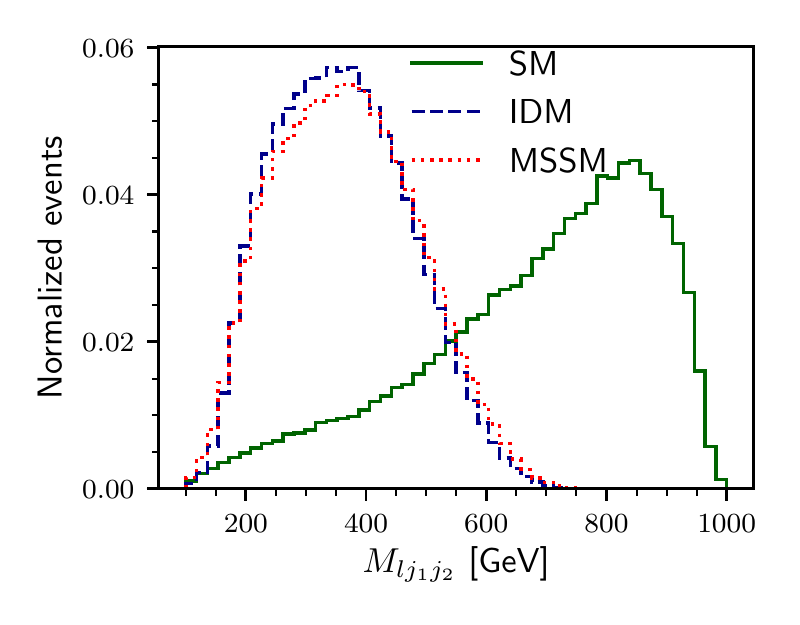}

	\caption{\label{fig:Lpol_kin-BP1} Normalized distributions of Kinematic variables in $e^- \, e^+ \rightarrow  l^\pm+2 j+\slashed{E}_T$  with $\sqrt{s}=1$ TeV and longitudinal polarized beams for {\tt BP1}.}
\end{figure} 
\begin{figure}[h!]
	\centering
	\includegraphics[width=0.45\textwidth]{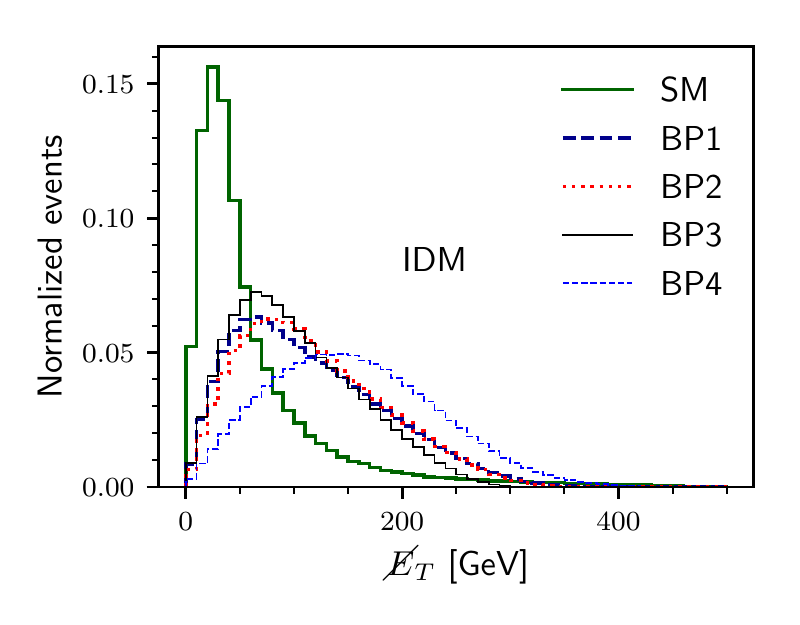}
	\includegraphics[width=0.45\textwidth]{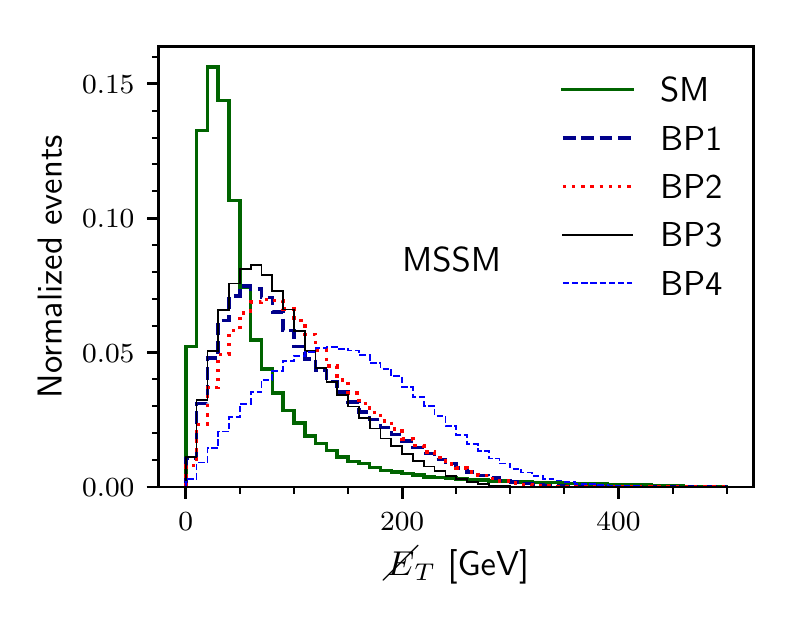}
	\caption{\label{fig:Lpol_kin-BPs} Normalized distribution of $\cancel{E}_T$ for all four {\tt BP}s of the IDM and MSSM 
	along with SM background in $e^- \, e^+ \rightarrow  l^\pm+2 j+\slashed{E}_T$  with $\sqrt{s}=1$ TeV and longitudinal polarized beams. }
\end{figure} 
The role of kinematic variables to analyse the signal against the SM background, which naturally arises in the final 
state $l^\pm2j+\cancel{E}_T$ are missing transverse energy ($\cancel{E}_T$), transverse momentum of lepton and 
jets ($p_{T_l}$, $p_{T_{j_1}}$, $p_{T_{j_2}}$), total visible mass ($M_{lj_1j_2}$), and energy of $jj$ pair which 
reconstructs the $W$ boson energy ($E_W$) and lepton ($E_l$). Normalized distribution of these kinematic 
variables are shown in Fig.~\ref{fig:Lpol_kin-BP1} for {\tt BP1} as representative with longitudinally 
polarized beams at $1$ TeV ILC set up for both IDM (in {\em dashed/blue} line) and MSSM (in {\em dotted/red} line) 
signal and SM background (in {\em solid/green} line). 

The $\cancel{E}_T$ distribution ({\em left-top} panel in Fig.~\ref{fig:Lpol_kin-BP1}) shows a peak in the lower energy 
values for the SM background compared to the signal. This is because the source of $\cancel{E}_T$ in the SM is
the neutrino coming from the $W$ boson decay, while for the signal, we have additional contributions coming from 
the undetected stable exotics  $C^0$  along with a neutrino.  Further, in the case of signal, the peak of $\slashed{E}_T$ 
shifts toward higher values as the splitting between the 
mass of the charged exotic and the stable neutral particle ($\Delta M$) increases, 
which can be seen in Fig.~\ref{fig:Lpol_kin-BPs} , where the SM background along with all four {\tt BP}s of 
IDM and MSSM  are shown. The peak for {\tt BP1} and {\tt BP3} with  $\Delta M=100$ GeV are at the same energy, while 
the peak for {\tt BP2} ($\Delta M=120$ GeV) and {\tt BP4}  ($\Delta M=290$ GeV) are at the higher 
energy side in  $\cancel{E}_T$ distribution due to more availability of energy for missing $C^0$  in the rest frame of 
it's charged partner.

We note that the two jets in the signal as well as the SM background come from the $W$ boson decay. 
Therefore it is possible to reconstruct the energy of the $W$ using the two jets. Ideally, for the 
SM background the $W$ bosons should be produced back to back with $E_W=\sqrt{s}/2=500$ GeV. However, 
the ISR effect gives an effective boost as well as changes the hard scattering collision energy, which along with 
hadronization and detector effects, result in the smearing of energies for the $W$ boson ($jj$ pair) and lepton as well as a longish tail
in the energy distribution of  $E_W$ extending to the kinematic threshold of $\sqrt{s_{eff}} \ge 2 M_W$ of the process. 
On the other hand, in the case of signal, the $W$ boson is produced from the decay of $C^\pm$, along with missing 
$C^0$. The available energy for $W$ is much lower than that of SM and is highlighted in the energy distribution for $E_W$. 
The difference between the endpoint of $E_W$ in SM and BSM models will be at least the $C^0$ mass. Similarly, 
the presence of missing $C^0$ can be seen from its imprint in other kinematic variables as well, given by  $E_l$, 
$M_{lj_1j_2}$ and $P_T$'s of jets and lepton, where the endpoint in  these distributions for the signal shifts towards the 
low energy values as compared to SM background. Hence,  rejection cuts near the endpoint of these variables 
help reduce the SM background significantly while keeping enough statistics for the signal.

We thus, implement a set of cuts on these kinematic variables as 
$\slashed{E}_T > 50 $ GeV, $P_{T_{l,j_1}} < 400$ GeV, $M_{l,j_1,j_2} < 700$ GeV, $E_l < 400$ GeV and $E_W < 500$ GeV 
to reduce the SM background while keeping enough statistics for the signal. We name these sets of cuts as {\tt Kin\_cut}.

\paragraph{\rm \bf Angular variables}
\begin{figure}[h!]
	\centering
	\includegraphics[width=0.45\textwidth]{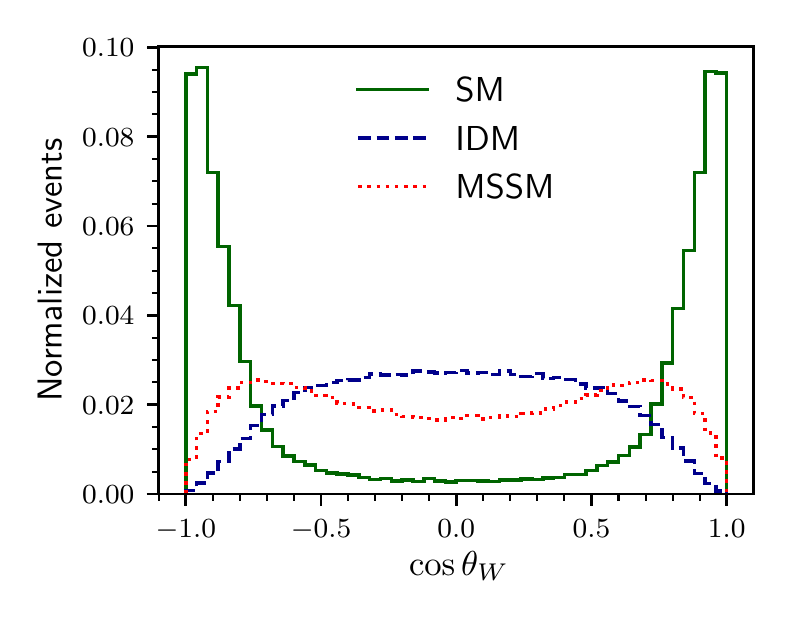}
	\includegraphics[width=0.45\textwidth]{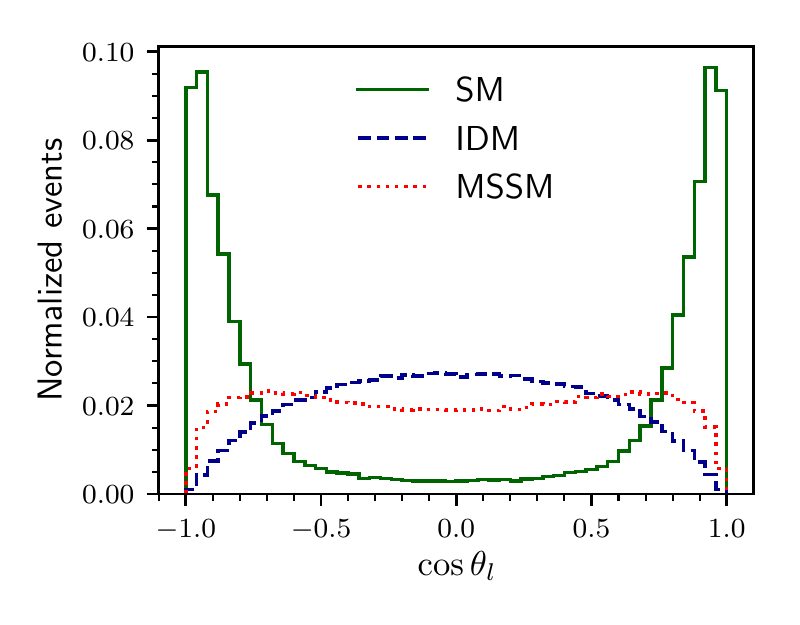}
	\includegraphics[width=0.45\textwidth]{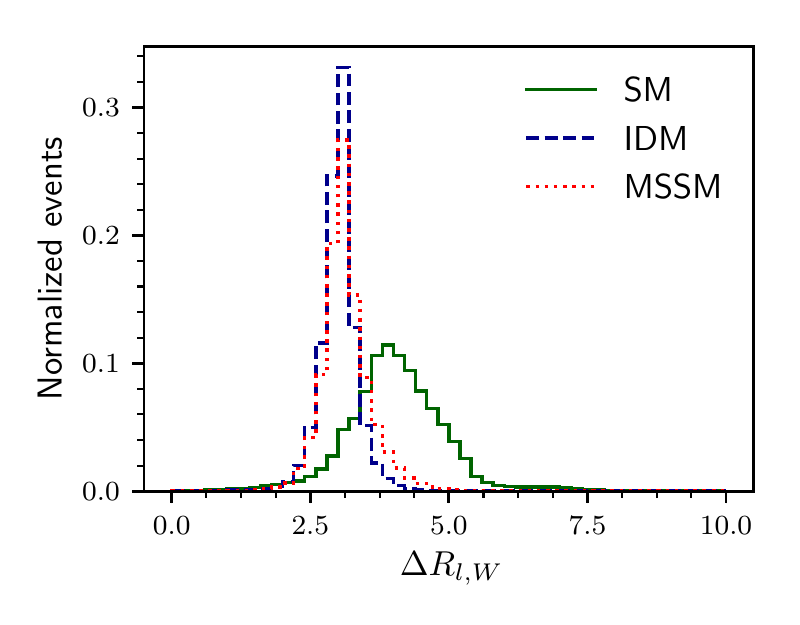}
	\caption{\label{fig:Lpol_ang-BP1} Normalized distribution of Angular variables  in $e^- \, e^+ \rightarrow  l^\pm+2 j+\slashed{E}_T$  with $\sqrt{s}=1$ TeV and longitudinal polarized beams for {\tt BP1}.}
\end{figure}
\begin{table}[h!]
	\centering
	\resizebox{16cm}{!}{
		\begin{tabular}{|c|c|c|c|c|c|c|c|c|c|}
			\cline{2-10}
			\multicolumn{1}{c|}{} 
			& SM background& \multicolumn{4}{c|}{IDM signal}&  \multicolumn{4}{c|}{MSSM signal } \\  \hline	
			Cuts  & $W^+W^-$  & ~~{\tt BP1}~~   & ~~{\tt BP2}~~ &  ~~{\tt BP3}~~ &  ~~{\tt BP4}~~ 
			&  ~~{\tt BP1}~~ & ~~{\tt BP2}~~ & ~~{\tt BP3}~~ &  ~~{\tt BP4}~~  \\ \hline		
			{\tt Sel\_cut}    & 22709.6   &558.29 & 462.28  &  456.62&321.47 & 5517.66  & 5337.85 &4948.52&5743.17 \\ \hline	
			{\tt Sel\_cut}+{\tt Kin\_cut}       & 3256.22  & 491.49 &420.57  &  400.05& 295.14 & 4738.46  & 4789.66 &4224.54&5153.8 \\ \hline
			{\tt Sel\_cut}+{\tt Kin\_cut} +{\tt Ang\_cut} & 1942.75 & 479.75   &  410.05  &  392.64& 274.70 & 4318.11  & 4352.4 & 3924.35 &4497.23\\  \hline	        	          \multicolumn{1}{c}{}&\multicolumn{1}{|c|}{Significance ($\mathcal{S}$)}  
			&10.47 & 9.00& 8.63&6.09  & 77.56& 78.08&71.55&80.25\\ 	\cline{2-10}
	\end{tabular}}
	\caption{ The cut-flow information on the $e^- \, e^+ \rightarrow  l^\pm +2 j+\slashed{E}_T$ process for both the signals and 
		background along with the significances for all four  {\tt BP}s  at the $1$ TeV ILC for $100$ fb$^{-1}$ of integrated luminosity.}
	\label{tab:cutflowLpol}
\end{table}   
It is quite well known that particle spin dictates the Lorentz structure of interaction vertices, and 
an efficient way of identifying these properties is through observing the kinematics in the angular 
variables of final state particles. For a scattering process such as $e^+e^- \to f \bar{f}$, where $f$ 
is a stable SM fermion; it is clearly highlighted in the polar distributions of the final state 
fermion. However, it becomes more challenging if the final state particles come from cascade decay 
of unstable particles produced as primaries in the $e^+e^-$ collisions, as in our case. We, therefore, explore the 
properties by looking at  the normalized distributions of some angular variables such as $\cos\theta$ of $W$ ($jj$ pair) 
and lepton and  the isolation variable $\Delta R = \sqrt{\Delta \eta^2 + \Delta \phi^2}$ between the lepton and $W$, 
which  are shown in Fig.~\ref{fig:Lpol_ang-BP1} for the SM background and signal (IDM and MSSM) for {\tt BP1} as 
representative. Owing to a $t$-channel sub-process in the SM process, $\cos\theta_{W}$ sharply peaks at $\pm 1$, i.e., along with the beam directions: The $W^-$ tends to remain toward $e^-$ direction, while the  $W^+$ tends to remain toward $e^+$ direction, see {\em left-top} panel in Fig.~\ref{fig:Lpol_ang-BP1}. On the other hand, there are only two $s$-channel sub-process in the IDM signal, which makes the $\cos\theta_{W}$ peak at $0$, i.e., transverse to the beam directions. In the case of the MSSM signal, there could be a $t$-channel sub-process due to sneutrino, which is kept heavy not to contribute to our analysis. Thus the $\cos\theta_{W}$ distribution has a small peak near $\pm 1$ with an overall flat shape throughout. The angular behavior of $W$ transfers to its decay product thus, lepton's $\cos\theta$ distribution is of similar nature to that of $W$, see {\em right-top}
panel. So an inclusion cut of $|\cos\theta| < 0.9$ for lepton and the $W$ will be effective to suppress the background with less effect on signals. In SM, the two intermediate $W$s are boosted in opposite directions, while in the case of signals, the $W$s are less boosted and also not in opposite directions as they are decayed from $C^\pm$ together with $C^0$s. Hence the $\Delta R _{l,W}$ attains higher values as compared to that of signals, see {\em bottom} panel in Fig.~\ref{fig:Lpol_ang-BP1}.
We, thus, put a rejection cut of $\Delta R_{lw} > 5.0$ to reduce SM background with less effect on the signals. We name the set of cuts on the angular variables as {\tt Ang\_cut}.

We estimate signal significance with an integrated luminosity of ${\cal L}=100$ fb$^{-1}$ with the formulae 
\begin{equation}\label{eq:significnae-formulae}
\mathcal{S} = \sqrt{2\left[\left(S+B\right)\log\left(1+\frac{S}{B}\right)-S\right]}
\end{equation}
with $S$ being the number of signal events and $B$ being the number of background events at a given luminosity. The signal significance of the IDM signal and MSSM signal are shown in Table~\ref{tab:cutflowLpol} for the successive cuts on kinematic variables and angular variables in all four {\tt BP}s. The significance values confirm that all four {\tt BP}s
are above $5~\sigma$ discovery limits in our analysis.

Although there is a good distinction between the background and the signals, it is ambiguous about the spin nature of particles contained in signals apart from the fact that the MSSM signal has more significance than the IDM signal with the same mass parameters. We try to estimate how the IDM signal is different from the MSSM signal based on the asymmetry of some angular variables in the following sub-section.

\subsubsection{\rm \bf Identifying the nature of signal}

We now try and identify the spin nature of the exotic charged particle ($C^+$) and its partner ($C^0$) by taking account of the 
observations made in the previous subsection. We try to identify the type of signal based on asymmetries constructed from 
the angular variables  $\cos\theta_{W}$ and  $\cos\theta_l$ for the signal inclusive of the background after effectively reducing 
the background contribution maximally as discussed earlier.  Then we estimate the difference between the signal from the two 
models  based on $\chi^2$ using the asymmetries. We have already discussed the difference in the shapes of 
$\cos\theta_{W/l}$ distribution, given in Fig.~\ref{fig:Lpol_ang-BP1} for the MSSM and IDM signal events. Based on the 
symmetric shape in $\cos\theta=0$, we define an asymmetry for both the variables as
\begin{equation}\label{eq:asymetry-def}
	\mathcal{A}\left( \cos\theta_{W/l} \right) = \frac{\sigma\left( |\cos\theta_{W/l}|>0.5\right) - \sigma\left(|\cos\theta_{W/l}| < 0.5\right)}{\sigma\left(|\cos\theta_{W/l}| > 0.5\right) + \sigma\left(|\cos\theta_{W/l}| < 0.5 \right)} .
\end{equation}
The difference is then estimated by a  $\chi^2$,  calculated as
\begin{equation}
\chi^2=\sum_i  \left|   \dfrac{\Delta{\cal A}^i = {\cal A}_{ \text{IDM+SM}}^i-{\cal A}_{ \text{MSSM+SM}}^i  }{\delta\mathcal{A}_{\text{\text{SM}}}^i }\right|^2,
\end{equation}
$i$ runs on the variables; $\delta\mathcal{A}_{\text{\text{SM}}} = \sqrt{\frac{{1-\mathcal{A}^2_{\text{\text{SM}}}}}{\sigma_{\text{\text{SM}}}\times\mathcal{L}}}$ is the statistical error on asymmetries due SM background. The asymmetries in each signal are estimated including the background after putting all the cuts ({\tt Sel\_cut}+{\tt Kin\_cut} +{\tt Ang\_cut}), i.e., reducing maximum background so as to highlight only the signal. 
The asymmetry of a signal ($S$) including background ($B$) will be given as,
 \begin{equation}\label{eq:Asb}
 \mathcal{A}_{S+B} = \frac{\mathcal{A}_S\sigma_S+\mathcal{A}_B\sigma_B}{\sigma_S+\sigma_B},
 \end{equation}
 i.e., weighted by their respective cross sections. 
The  estimated values of asymmetries of the variables in each signal mixed with the background are shown in Table~\ref{tab:AsymLpol} (top two rows).
\begin{table}[h]
	\centering
	\resizebox{16cm}{!}{
		\begin{tabular}{|c|c|c|c|c|c|c|c|c|c|}
			\cline{2-10}
			\multicolumn{1}{c|}{} 
			& \multicolumn{1}{c|}{$\mathcal{A}_{\text{SM}}$}& \multicolumn{4}{c|}{$\mathcal{A}_{\text{SM+IDM}}$}&  \multicolumn{4}{c|}{$\mathcal{A}_{\text{SM+MSSM}}$ } \\  \hline	
			Variables & ~~$W^{+}W^{-}$~~ & ~~{\tt BP1}~~   & ~~{\tt BP2}~~ &  ~~{\tt BP3}~~ &  ~~{\tt BP4}~~ 
			&  ~~{\tt BP1}~~ & ~~{\tt BP2}~~ & ~~{\tt BP3}~~ &  ~~{\tt BP4}~~  \\ \hline		
			$\cos\theta_W$    & 0.1682 &  0.0632619& 0.0663791& 0.0573263&0.118111& 0.0260318& -0.030412& -0.0775697& -0.00320661 \\ \hline
			$\cos\theta_l$    & 0.291 &  0.169782& 0.182949& 0.180954&0.225528& 0.0212554& -0.0130694& -0.0379408& 0.0206557 \\ \hline \hline       
			&$\delta\mathcal{A}_{\text{SM}}$ &  \multicolumn{8}{c|}{$\left|\frac{\Delta\mathcal{A}}{\delta\mathcal{A}_{\text{SM}}}\right|$ }\\ \hline
			& ~~$W^{+}W^{-}$~~     & \multicolumn{2}{c|}{~~{\tt BP1}~~} & \multicolumn{2}{c|}{~~{\tt BP2}~~} & \multicolumn{2}{c|}{~~{\tt BP3}~~} &  \multicolumn{2}{c|}{~~{\tt BP4}~~}\\ \hline	
		$\cos\theta_W$    &0.0223  &    \multicolumn{2}{c|}{1.66469}&\multicolumn{2}{c|}{4.32789}&\multicolumn{2}{c|}{ 6.0317}&\multicolumn{2}{c|}{5.42458} \\ \hline
			$\cos\theta_l$    & 0.0217     & \multicolumn{2}{c|}{6.84267} & \multicolumn{2}{c|}{9.03066} & \multicolumn{2}{c|}{10.0846} & \multicolumn{2}{c|} {9.43857} \\ \hline
			\multicolumn{2}{|c|}{$\chi^{2}$}  & \multicolumn{2}{c|}{49.5933} & \multicolumn{2}{c|}{100.2823} &\multicolumn{2}{c|}{138.0805}&\multicolumn{2}{c|}{118.5126} \\ \hline 
	\end{tabular}}
	\caption{ Asymmetries for both the signals and background and difference between the signals  in the 
	$e^- \, e^+ \rightarrow  \ell+2 j+\slashed{E}_T$ process  for longitudinally polarized beams for 
	${\cal L}=100$ fb$^{-1}$ of integrated luminosity. }
	\label{tab:AsymLpol}
\end{table} 
Although the  ${\cal A}_S$ for $\cos\theta_{W/l}$ seems to be $(-)$ve for IDM signal and $(+)$ve for MSSM signal by looking Fig.~\ref{fig:Lpol_ang-BP1}, the ${\cal A}_{S+B}$ are $(+)$ve for IDM, while they are $(-)$ve for MSSM in most of the {\tt BP}. The reasons are the following.  The asymmetries in SM are large $(+)$ve, and the SM  cross section is larger by roughly a factor of $5$ compared to the cross sections in IDM after all cuts, see Table~\ref{tab:cutflowLpol}. Hence the  asymmetries in IDM signal reduces but remain positive when combined with the SM background weighted by cross sections in accordance with Eq.~(\ref{eq:Asb}). For the case of MSSM, individual asymmetries in signal become $(-)$ve after the inclusion cut $|\cos\theta_{W/l}| < 0.9$. Moreover, the  cross sections are larger in MSSM compared to SM by roughly a factor of $2$, see Table~\ref{tab:cutflowLpol}. This makes the ${\cal A}_{S+B}$ to be $(-)$ve for most of the {\tt BP}. 
The changes in the asymmetries w.r.t. background are above $1\sigma$ statistical errors shown in the second column in Table~\ref{tab:AsymLpol}. Differences between the signals are calculated in terms of $\chi^2$ and shown in Table~\ref{tab:AsymLpol} in the last three rows for all {\tt BP}s for an integrated luminosity of ${\cal L}=100$ fb$^{-1}$. The two signals have a difference of  $\sim 10\sigma$ in all the four {\tt BP}s based on the total $\chi^2$ shown in the last row.

These asymmetries can help to identify the nature of the signal if observed above the background. The asymmetries remain $(+)$ve for IDM, while they become $(-)$ve and above statistical errors for {\tt BP2} and {\tt BP3} due to having large $M_0$, and $(+)$ve and within $1\sigma$ error for other {\tt BP}s. Nevertheless,   the characteristics of the asymmetries which can be useful to distinguish between the signals are benchmark dependent. We require a variable that can distinguish between the signals irrespective of the benchmark masses; we will explore such a possibility in the following section by using transverse beam polarizations.

\subsection{Analysis with transverse beam polarization}\label{sec:Tpol-analysis}
\begin{figure}[th!]
	\centering
	\includegraphics[width=0.5\textwidth]{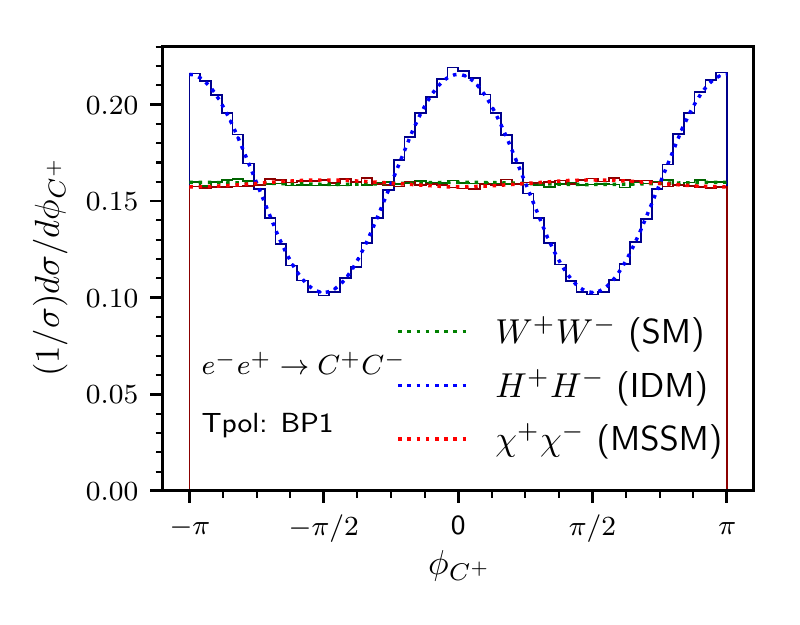}	\caption{\label{fig:Tpol-BP1-phiC} Normalized distribution of the azimuthal angle $\phi$ of the charged particle ($C^+$) at production level  for the background and the signals in {\tt BP1} in $e^-e^+$ collider with transversely polarized beams at $\sqrt{s}=1$ TeV.}
\end{figure}  
We observe in the previous section that though the angular variables $\cos\theta_{W}$ and $\cos\theta_l$ are 
useful in identifying the nature of the signal, they are limited to being dependent on benchmark selections. The transversely 
polarized beams have the potential of exploring new physics from the azimuthal distributions of final state 
particles~\cite{Ananthanarayan:2004xf,Burgess:1990ba,Rizzo:2002ww,Ananthanarayan:2004eb,Rindani:2004ue,Hikasa:1985qi,Godbole:2009qv}. 
Here, we use the facility of transverse polarization of $e^\pm$ beams to showcase the distribution of azimuthal 
angle $\phi$, which has a very distinctive nature for two signals irrespective of the benchmark selections as 
well as the background. Before looking at the distribution of $\phi$ for the final states, we first observe 
the distribution at the production level and make an ansatz at the final state particle. We calculate the normalized 
$\phi$ distribution for the $C^+$ analytically in $e^-e^+\to C^+C^-$ process for SM as well as for the signals 
with transverse beam polarization of  ($80 \%$, $60 \%$) for ($e^-$, $e^+$) beams and show them in 
Fig.~\ref{fig:Tpol-BP1-phiC} for {\tt BP1} with {\em dotted} lines. The same distributions for $\phi_C$ are 
also shown with {\em solid} lines computed from events generated in 
{\tt WHIZARD}\footnote{ISR effect has been neglected here for simplicity.}.  
The differential cross section in $\phi_C$  has the form
\begin{equation}\label{eq:gen-exprsn-phi}
\frac{d\sigma}{d\phi_C} = \frac{\sigma}{2\pi} + \eta_T\xi_T f(\beta_C) \cos\left( 2\phi_C-\phi_{e^-}+\phi_{e^+}\right),
\end{equation}
where $f(\beta_C)$ is a function of boost ($\beta$) for the particle $C$ with different form for different physics model, see appendix~\ref{app:tpol-analytics} for details. We choose the spin direction for both the initial beams along the $(+)$ve $x$-axis as an example; making the spin directions opposite for the two beams would result in an overall phase-shift for all models not affecting our findings. The explicit form of $f(\beta_C)$ ({Shown in Eqs.~ (\ref{eq:phi-sm}), (\ref{eq:phi-idm}) and (\ref{eq:phi-mssm}) for analytical expressions}) makes the amplitude large for the IDM signal, while negligibly small for the MSSM signal and the $W^+W^-$ background, see Fig.~\ref{fig:Tpol-BP1-phiC}. If the $\phi_C$ distribution in MSSM attains an amplitude comparative to IDM in some other benchmark scenario, the nature will be different due to having a relative $\pi$ phase shift compared to IDM, as can be seen in  Fig.~\ref{fig:Tpol-BP1-phiC}.
\begin{figure}[t!]
	\centering
	\includegraphics[width=0.45\textwidth]{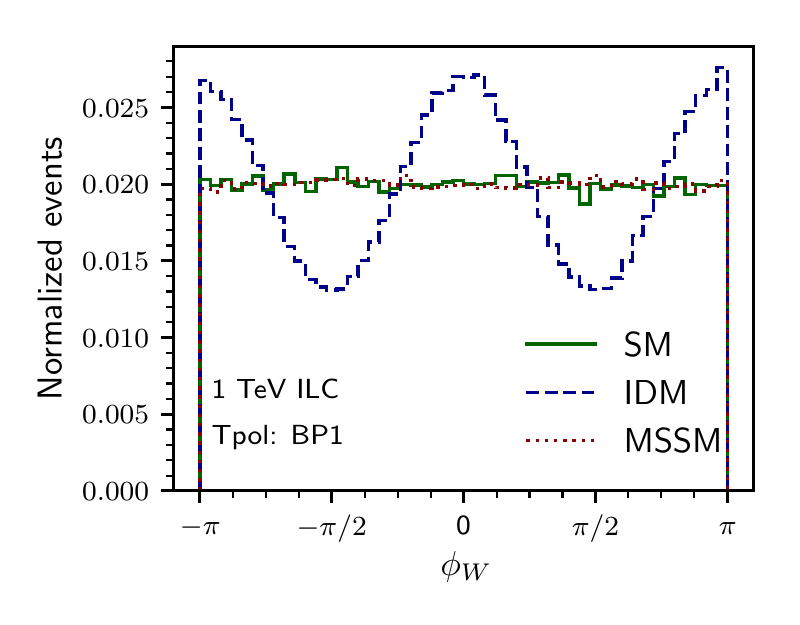}
	\includegraphics[width=0.45\textwidth]{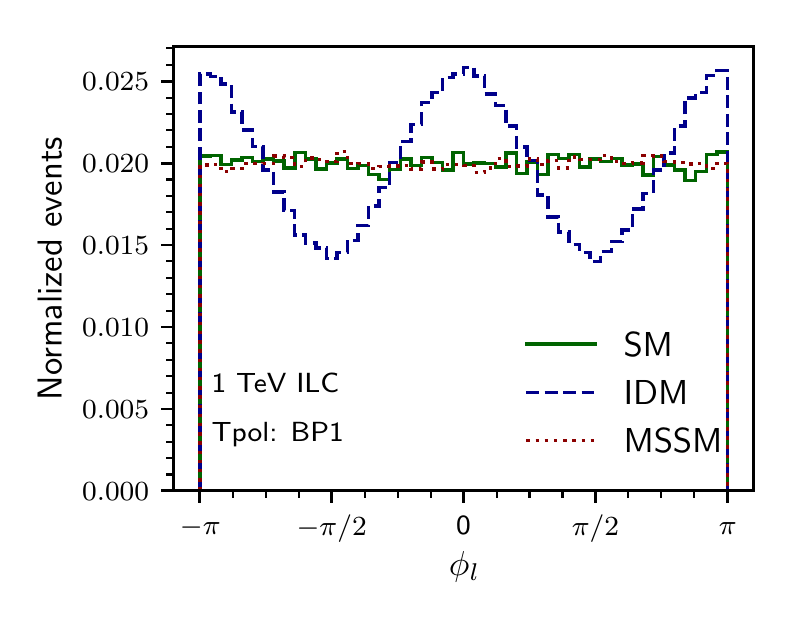}
	\caption{\label{fig:Tpol-BP1-phiWl} Normalized distribution of $\phi_W$  ({\em left-panel} ) and $\phi_l$ ({\em right-panel}) for the background and the signals in {\tt BP1} in $e^+e^-$ collider for $\sqrt{s}=1$ TeV with transversely polarized beams.}
\end{figure}
\begin{table}[h!]
	\centering
	\resizebox{16cm}{!}{
		\begin{tabular}{|c|c|c|c|c|c|c|c|c|c|c|}
			\cline{2-11}
			\multicolumn{1}{c|}{} 
			& $\mathcal{A}_{\text{SM}}$&$\delta\mathcal{A}_{\text{SM}}$  &\multicolumn{4}{c|}{$\mathcal{A}_{\text{SM+IDM}}$}&  \multicolumn{4}{c|}{$\mathcal{A}_{\text{SM+MSSM}}$ } \\  \hline	
			Variables  & & & ~~{\tt BP1}~~   & ~~{\tt BP2}~~ &  ~~{\tt BP3}~~ &  ~~{\tt BP4}~~ 
			&  ~~{\tt BP1}~~ & ~~{\tt BP2}~~ & ~~{\tt BP3}~~ &  ~~{\tt BP4}~~  \\ \hline		
			$\phi_W$    & -0.0029 &0.0379   &0.0471 & 0.0382  & 0.0387 &0.0116   & -0.0053 &-0.0032&-0.0057& -0.0038\\ \hline	
			$\phi_l$    & 0.0113 & 0.0379  & 0.0475 & 0.0378 &0.0375 &  0.0222 & -0.0018  & 0.0009 &0.0001&0.0027 \\ \hline		 
	\end{tabular}}

	\caption{ Azimuthal asymmetries for both the signals and background  in the $e^- \, e^+ \rightarrow  \ell+2 j+\slashed{E}_T$ 
	process  for transversely polarized initial beams. The error in the SM is shown  for ${\cal L}=100$ fb$^{-1}$ of integrated 
	luminosity.}
	\label{tab:AsymTpol}
\end{table} 
This behavior of $\phi$ distribution for the mother particle $C^+$ is  transferred to its decay products which in this case are the 
lepton and $W(jj)$ even after including the detector level effects as well as with ISR effect, as seen 
in Fig.~\ref{fig:Tpol-BP1-phiWl}. Here also, the amplitude in IDM is large enough compared to MSSM 
and SM, making the $\phi_{W/l}$ a good discriminator for identifying the nature of signal. To depict a quantitative measure of distinction between the two signals, we again construct asymmetries for the  $\phi_{W/l}$ given by
\begin{equation}\label{eq:asym-phi}
\mathcal{A}\left(\phi_{W/l}\right) = \frac{\sigma\left(\cos\left(2\phi_{W/l}\right) > 0\right) 
	- \sigma\left(\cos\left(2\phi_{W/l}\right) < 0\right)}{\sigma\left(\cos\left(2\phi_{W/l}\right) > 0\right) 
	+ \sigma\left(\cos\left(2\phi_{W/l}\right) < 0\right)} ~.
\end{equation}     
We have listed the asymmetry values  in Table~\ref{tab:AsymTpol} 
for the SM background as well as all four {\tt BP}s combined with the background after the
cuts {\tt Sel\_cut}+{\tt Kin\_cut} +{\tt Ang\_cut}, as used in the previous section for our analysis with longitudinally polarized beams.
The asymmetries for both $\phi_W$ and $\phi_l$ are nearly zero (much smaller than the $1\sigma$ statistical error given in the first and second rows of the second column) for the background (shown in the second column) and the MSSM signal. The asymmetries for the IDM combined with the background, however, remain large enough even after mixing with the background with a larger cross section compared to the signal.
Although the asymmetry in IDM reduces as the $m_H$ increases ({see Eq. (\ref{eq:phi-idm}) for phi analytic}) 
mainly due to smaller cross sections, asymmetry can still be significant enough with higher integrated luminosity.  
Thus these two distributions and their asymmetries are able to identify the spin nature of $C^+$/$C^0$
contained in the signal if a significant deviation is observed at future ILC. The signal identification can be quantified with 
higher significance for higher luminosity. 

\section{Conclusion}\label{sec:conclusion}
In this article, we make an effort to identify the spin nature of exotic charged particles and their neutral partner, possibly 
a dark matter candidate, in  $l^\pm2j+\cancel{E}_T$ final state at the future $e^+e^-$ collider. We chose two well known 
BSM models such as IDM and MSSM, containing exotic charged particles and  neutral partners of spin zero (scalar) 
and half (fermion), respectively, for the demonstration. First, we devised some rectangular cuts on some kinematic 
as well as angular variables to reduce the SM background in order to find a significant deviation from the SM 
background. We then use the shape of angular variables and their asymmetries to identify the nature of the 
new physics signals. 
The $\cos\theta_{W/l}$ distribution and their asymmetries help distinguish and identify the MSSM and the IDM signals over the SM backgrounds with longitudinally polarized beams depending on the benchmark points.
The transverse beam polarization helps to identify the IDM signal robustly through the azimuthal ($\phi_{W/l}$) 
distribution of the final state particles. Thus both longitudinal polarization and transverse polarization 
prove useful in their respective ways in identifying the spin nature of exotic charged particles and their partner 
contained in a new physics signal if observed at the future $e^-e^+$ collider in $l^\pm2j+\cancel{E}_T$ final state. 
We note that while the identification gets help from the enhanced cross section in the longitudinal polarization 
due to the large cross section, a more clear and more distinct identification is established with the use of transversely 
polarized beam.   

\section*{Acknowledgments}

The authors would like to acknowledge the support from DAE, India for the Regional Centre for Accelerator 
based Particle Physics (RECAPP), Harish Chandra Research Institute.

	\appendix
\section{Feynman rules}
The Feynman rules for the vertices involved in the SM, IDM and MSSM processes are 
\begin{itemize}
	\item $H^+(k_1)H^-(k_2)A_{\mu}$ : \hspace*{0.8cm} $ -i\,  e\, (k_2 - {k_1})_{\mu}$,
	\item $H^+(k_1)H^-(k_2)Z_{\mu}$ : \hspace*{0.8cm} $-i\,  (- g_1 \sin\theta_W + g_2 \cos\theta_W)\, (k_2 - {k_1})_{\mu}$,
	\item $\chi^+_1\chi^-_1A_{\mu}$ : \hspace*{2.4cm} $ -i\,  e\, \gamma_{\mu}$,
	\item $ \chi^+_1\chi^-_1 Z_{\mu}$ : \hspace*{2.7cm} $ i\,   \gamma_{\mu} (v_\chi - a_\chi \gamma_5) = i\,  g_Z\, \gamma_{\mu} \left((g_L + g_R) +(g_R - g_L) \,\gamma_5\right) $, 
\end{itemize}
where		$g_L = \left(\sin^2{\theta_W} - \frac{3}{4} - \frac{1}{4}(2 \cos^2{\theta_L - 1})\right),~
g_R = \left(\sin^2{\theta_W} - \frac{3}{4} - \frac{1}{4}(2 \cos^2{\theta_R - 1})\right)$,
and $\cos{\theta_L} = \mathcal{U}_{11}$, $\cos{\theta_R} = \mathcal{V}_{11}$ are the cosine of left and right chargino mixing angles. 

\section{Azimuthal angular distribution with transversely polarized beams }\label{app:tpol-analytics}	
\subsubsection{\rm\bf SM: $ e^-(p_1,s_1) e^+(p_2,s_2) \to W^-(k_2,\lambda_2) W^+(k_1,\lambda_1)$} 
The amplitude for the diagram with the photon in the $s$-channel is given by
\begin{align}
\mathcal{M}_{\gamma}(s_1,s_2,\lambda_1,\lambda_2) = - \frac{\,i e^2 }{s} \,  \bar{v}(s_2,p_2)  \gamma^{\mu} u(s_1,p_1) \epsilon^{\nu}(\lambda_1,k_1){\epsilon^{\rho}}^{*}(\lambda_2,k_2) G_{\mu \nu \rho}(k_1,k_2),
\end{align}
where
\begin{equation}
G_{\mu \nu \rho}(k_1,k_2) \equiv ((k_2 - {k_1})_{\mu}g_{\nu \rho}-(2k_2+k_1)_\nu g_{\rho \nu}+ (2k_1+k_2)_\rho g_{\mu \nu}).
\end{equation}
Following are the expressions for the amplitudes of the diagrams containing $Z$ boson and neutrino propagator, respectively, 
\begin{eqnarray}
\mathcal{M}_{Z}(s_1,s_2,\lambda_1,\lambda_2) =& \frac{i e^2}{2\,(s-M^2_Z)\, \sin^2{\theta_W} }  \,  \bar{v}(s_2,p_2)  \gamma^{\mu} (v_f - a_f \gamma_5) u(s_1,p_1) \nonumber\\
&\times\epsilon^{\nu}(\lambda_1,k_1){\epsilon^{\rho}}^{*}(\lambda_2,k_2) G_{\mu \nu \rho}(k_1,k_2),  
\end{eqnarray}
\begin{equation}
\mathcal{M}_{\nu}(s_1,s_2,\lambda_1,\lambda_2) = - \frac{i {g_2}^2  }{(p_1-k_1)^2} \,  \bar{v}(s_2,p_2)  \gamma^{\nu} P_L (\slashed{p}_1-\slashed{k}_1) \gamma^{\mu} P_L u(s_1,p_1) \epsilon^{\mu}(\lambda_1,k_1){\epsilon^{\nu}}^{*}(\lambda_2,k_2). 
\end{equation}
The square amplitude with polarized initial beams is given by
\begin{equation}
|\mathcal{M}|^2 = \sum_{s_i,\lambda_j}\mathcal{M} (s_1,s_2,\lambda_1,\lambda_2) \mathcal{P}_{e^{-}}(s_1,s_3) \mathcal{P}_{e^{+}}(s_2,s_4) \mathcal{M}^{\dagger} (s_3,s_4,\lambda_1,\lambda_2), 
\end{equation}
where
\begin{equation}
\mathcal{M}  = \mathcal{M}_{\gamma} + \mathcal{M}_{Z} + \mathcal{M}_{\nu}.
\end{equation}
The spin density matrix for electron and positron ($\mathcal{P}_{e^{-}}$,$\mathcal{P}_{e^{+}}$) are written as
\begin{equation}
\mathcal{P}_{e^{-}} = \left(
\begin{array}{cc}
\frac{1}{2} & \frac{1}{2} e^{-i \phi_{e^{-}} } \eta_T  \\
\frac{1}{2} e^{i \phi_{e^{-}} } \eta_T  & \frac{1}{2} \\
\end{array}
\right),
\end{equation}
\begin{equation}
\mathcal{P}_{e^{+}} = \left(
\begin{array}{cc}
\frac{1}{2} & \frac{1}{2} e^{-i \phi_{e^{+}} } \xi_T  \\
\frac{1}{2} e^{i \phi_{e^{+}} } \xi_T  & \frac{1}{2} \\
\end{array}
\right)
\end{equation}
with their degree of polarization along transverse direction ($\theta = \frac{\pi}{2}$, $\phi_{e^{-}/e^{+}}$) as ($\eta_T$, $\xi_T $), respectively, 

Throughout the calculation we use $c\theta $ and $s\theta $ as shorthand notations for $\cos{\theta}$ and $\sin{\theta}$ respectively. The total square amplitude can be written as 
\begin{equation}
| \mathcal{M}|^2 = |\mathcal{M}_\gamma|^2  +|\mathcal{M}_Z|^2 +|\mathcal{M}_\nu|^2
+2Re\left(\mathcal{M}_\gamma\mathcal{M}_Z^\dagger\right) +2Re\left(\mathcal{M}_\gamma\mathcal{M}_\nu^\dagger\right)
+2Re\left(\mathcal{M}_\nu\mathcal{M}_Z^\dagger\right).
\end{equation}	
The expressions for the square amplitude of various diagrams and  their interference are given by

\begin{eqnarray}
|\mathcal{M}_\gamma|^2     & = &\frac{1}{2 \left(\beta ^2-1\right)^2} \Bigg[\beta ^2  e^4  \left(3 \beta ^4-18 \beta ^2-\left(3 \beta ^4-2 \beta ^2+3\right) c\theta ^2+19\right) \nonumber \\
&- &\beta ^2 \left(3 \beta ^4-2 \beta ^2+3\right) \left(c\theta ^2-1\right)  e^4 \eta_T  \xi_T   \cos (2 \phi )\cos \left(\phi_{e^{-}} -\phi_{e^{+}} \right) \nonumber \\
&-&   \beta ^2 \left(3 \beta ^4-2 \beta ^2+3\right) \left(c\theta ^2-1\right)  e^4 \eta_T  \xi_T   \sin (2 \phi ) \sin \left(\phi_{e^{-}} -\phi_{e^{+}} \right)\Bigg], 
\end{eqnarray}
\begin{eqnarray}
|\mathcal{M}_Z|^2 & = & \frac{2 \beta ^2 E_e^4 e^4}{\sin^4{\theta_W}\left(\beta ^2-1\right)^2 \left(s-M^2_Z\right)^2}\Bigg[ \left(a_f ^2+v_f ^2\right) \left(3 \beta ^4-18 \beta ^2-\left(3 \beta ^4-2 \beta ^2+3\right) c\theta ^2+19\right) \nonumber \\
& & +  \left(3 \beta ^4-2 \beta ^2+3\right) \left(c\theta ^2-1\right)  \eta_T   \xi_T  \left(a_f ^2-v_f ^2\right)
\cos (2 \phi ) \cos \left(\phi_{e^{-}} -\phi_{e^{+}} \right) \nonumber \\
& & -  \left(3 \beta ^4-2 \beta ^2+3\right) \left(c\theta ^2-1\right)  \eta_T   \xi_T 
\left(a_f ^2-v_f ^2\right) \sin (2 \phi )\sin \left(\phi_{e^{-}} -\phi_{e^{+}} \right)\Bigg],  \nonumber\\
\end{eqnarray}

\begin{eqnarray}
|\mathcal{M}_\nu|^2 &=& \frac{e^4}{4\sin^4\theta_W \left(\beta ^4-2 \beta ^3 c\theta +2 \beta  c\theta -1\right)^2}\Bigg[  \left(\beta ^6-10 \beta ^4+9 \beta ^2-4 \beta ^4 c\theta ^4+4 \left(\beta ^5+\beta ^3\right) c\theta ^3\right)\nonumber \\
&-&  \left(\left(5 \beta ^6+6 \beta ^4-11 \beta
^2\right) c\theta ^2+4 \left(3 \beta ^4-\beta ^2-4\right) \beta  c\theta +4\right)\Bigg],
\end{eqnarray}

\begin{eqnarray}
Re\left(\mathcal{M}_\gamma\mathcal{M}_Z^\dagger\right) & =  & \dfrac{2 \beta ^2 E_e^2  e^4}{2\sin^2{\theta_W}\left(\beta ^2-1\right)^2 \left(s-M^2_Z\right)}\Bigg[   v_f  \left(18 \beta ^2-3 \beta ^4+\left(3 \beta ^4-2 \beta ^2+3\right) c\theta
^2-19\right) \nonumber \\
&   + & \left(3 \beta ^4-2 \beta ^2+3\right) \left(c\theta ^2-1\right)  \eta_T   \xi_T  
v_f  \cos (2 \phi ) \cos \left(\phi_{e^{-}} -\phi_{e^{+}} \right) \nonumber \\
&	+ & \left(3 \beta ^4-2 \beta ^2+3\right) \left(c\theta ^2-1\right)  \eta_T   \xi_T  
v_f  \sin (2 \phi ) \sin \left(\phi_{e^{-}} -\phi_{e^{+}} \right)\Bigg],\nonumber\\
\end{eqnarray}
\begin{eqnarray}
Re\left(\mathcal{M}_\gamma\mathcal{M}_\nu^\dagger\right) & = & \dfrac{\beta ^2  e^4 }{4 \sin^2{\theta_W} \left(\beta ^2-1\right)^2 \left(2 \beta  c\theta -\beta ^2-1\right)}\Bigg[ \left(2 \beta ^2-3 \beta
^4-3  \right) c\theta ^2 \nonumber \\
&+& \left(3 \beta ^4-18 \beta ^2+19\right)+2 \left(\beta ^3+\beta ^1\right) c\theta ^3 \nonumber \\		
& + & \beta ^2  e^4 \eta_T   \xi_T   \left(2 \left(\beta ^3+\beta \right) c\theta
^3+\left(3 \beta ^4-2 \beta ^2+3\right) s\theta ^2\right) \cos (2 \phi ) \cos \left(\phi_{e^{-}} -\phi_{e^{+}} \right)  \nonumber \\
& + &   \eta_T  \xi_T    \left(2 \left(\beta ^3+\beta \right) c\theta
^3+\left(3 \beta ^4-2 \beta ^2+3\right) s\theta ^2\right) \sin (2 \phi ) \sin \left(\phi_{e^{-}} -\phi_{e^{+}} \right) \nonumber\\
&-&  \eta_T   \xi_T   \left(2 \left(\beta ^3+\beta \right) c\theta \right) \cos (2 \phi ) \cos \left(\phi_{e^{-}} -\phi_{e^{+}} \right)+\left(6 \beta ^4-2 \beta ^2-8\right) c\theta   \nonumber \\
&-& \eta_T   \xi_T   \left(2 \left(\beta ^3+\beta \right) c\theta \right) \sin (2 \phi ) \sin \left(\phi_{e^{-}} -\phi_{e^{+}} \right)\Bigg],	
\end{eqnarray}

\begin{eqnarray}
Re\left(\mathcal{M}_\nu\mathcal{M}_Z^\dagger\right)& = & \dfrac{\beta  E_e^2 e^4}{2\sin^4{\theta_W} \left(\beta ^2-1\right)^2 \left(2 \beta  c\theta -\beta ^2-1\right)\left(s-M^2_Z\right)}\Bigg[  \beta(a_f+v_f) \left(2 \left(\beta ^3+\beta \right)
c\theta ^3\right) \nonumber \\
&+&\beta(a_f+v_f)	3 \beta ^4s\theta ^2-  \left(18 \beta ^2-19\right)\nonumber \\
&	 +& \eta_T    \xi_T   (a_f-v_f) \left(\left(3 \beta ^4-2 \beta ^2+3\right) s\theta ^2\right) \cos (2 \phi ) \cos \left(\phi_{e^{-}} -\phi_{e^{+}} \right) \nonumber \\
&  + &  \beta \eta_T    \xi_T   (a_f-v_f) \left(\left(3 \beta ^4-2 \beta ^2+3\right) s\theta ^2\right) \sin (2 \phi ) \sin \left(\phi_{e^{-}} -\phi_{e^{+}} \right) \nonumber \\
&+&    (a_f+v_f) \left(\left(2 \beta ^3-3 \beta \right) c\theta ^2+\left(6 \beta ^4-2 \beta ^2-8\right) c\theta \right)\nonumber \\
&-& \beta \eta_T   \xi_T   \left(2 \left(\beta ^3+\beta \right) c\theta s\theta ^2\right) \sin (2 \phi ) \sin \left(\phi_{e^{-}} -\phi_{e^{+}} \right) \nonumber \\
&-& \beta\eta_T   \xi_T   \left(2 \left(\beta ^3+\beta \right) c\theta s\theta ^2\right) \cos (2 \phi ) \cos \left(\phi_{e^{-}} -\phi_{e^{+}} \right) \Bigg].
\end{eqnarray}
Here, $\beta$ is the boost of the $W$ bosons and $\sqrt{s}=2 E_e$ is the center of mass energy.
Integrating over the polar angle $\theta$, we get the differential cross section with respect to the azimuthal angle ($\phi$) as
\begin{equation}\label{eq:dsig-dphi}
\frac{d\sigma}{d\phi} =  \dfrac{\beta}{64\pi^2 s} \int d\cos\theta~ | \mathcal{M}|^2 \times \left(3.894\times 10^{11}\right)~~\text{fb}.
\end{equation}
In this case,
\begin{eqnarray}
\frac{d\sigma}{d\phi}& = &\frac{\beta \times 3.894 \times 10^{11}}{64 \pi^2 s}\Bigg(\frac{E_e^2 e^4   (a_f+v_f) \left(4 \beta  \left(3 \beta ^6-23 \beta ^4+\beta ^2+27\right)\right)}{12 \beta  M^4_W \left(s-M^2_Z\right)}\nonumber \\
&+&\frac{E_e^2 \eta_T  e^4  \xi_T   (a_f-v_f) \left(3
	\left(\beta ^2-1\right)^4 \log \left(\frac{1+\beta }{1-\beta }\right)^2\right) \cos (2 \phi -\phi_{e^{-}} +\phi_{e^{+}} )}{12 \beta  M^4_W
	\left(s-M^2_Z\right)\sin^4{\theta_W}}\nonumber \\
&+&\frac{16 \beta ^3  e^4  \left(3 \beta ^4-26 \beta ^2+\left(3 \beta ^4-2 \beta ^2+3\right) \eta_T  \xi_T  \cos (2 \phi -\phi
	_e+\phi_ p)+27\right)}{24 \beta 
	M^4_W}\nonumber \\
& + &\frac{8 \beta ^2 E_e^4 e^4  \left(\left(3 \beta ^4-26 \beta ^2+27\right) \left(a_f^2+v_f^2\right)\right)}{3 M^4_W \left(s-M^2_Z\right)^2\sin^4{\theta_W}} \nonumber \\
& - &\frac{8 \beta ^2 E_e^2  e^4  v_f \left(3 \beta ^4-26 \beta ^2+\left(3 \beta ^4-2 \beta ^2+3\right) \eta_T  \xi_T 
	\cos (2 \phi -\phi_ e+\phi_ p)+27\right)}{3 M^4_W \left(s-M^2_Z\right)\sin^2{\theta_W}} \nonumber \\
& + &\frac{ e^4 \eta_T   \xi_T   \left(3 \left(\beta ^2-1\right)^4
	\log \left(\frac{1+\beta }{1-\beta}\right)^2-4 \beta  \left(3 \beta ^6+\beta ^4+\beta ^2+3\right)\right) \cos (2 \phi -\phi_ e+\phi_ p)}{24 \beta  M^4_W \sin^2{\theta_W}} \nonumber \\
&-&\frac{ e^4  \left(3 \beta ^5-\beta ^3+3 \left(\beta ^2-1\right)^2 \left(\beta ^2+1\right) \tanh ^{-1}(\beta )-3 \beta \right)}{3 \beta 
	M^4_W \sin^4{\theta_W}}-\frac{e^4\left(4 \beta  \left(3 \beta ^6-23 \beta
	^4+\beta ^2+27\right)\right)}{24 \beta  M^4_W} \nonumber\\
&-&\frac{e^4\left(3 \left(\beta ^2-9\right) \left(\beta ^2-1\right)^3 \log \left((\beta -1)^2\right)-6 \left(\beta ^2-9\right) \left(\beta ^2-1\right)^3 \log (\beta +1)\right)}{24 \beta  M^4_W} \nonumber \\
&-&\frac{E_e^2 e^4   (a_f+v_f) \left(3 \left(\beta
	^2-9\right) \left(\beta ^2-1\right)^3 \log \left(\frac{1+\beta}{1-\beta }\right)^2\right)}{12 \beta  M^4_W \left(s-M^2_Z\right)}\nonumber \\
&-&\frac{E_e^2 \eta_T  e^4  \xi_T   (a_f-v_f) \left(4   \left(3 \beta ^6+\beta ^4+\beta ^2+3\right)\right) \cos (2 \phi -\phi_{e^{-}} +\phi_{e^{+}} )}{12   M^4_W
	\left(s-M^2_Z\right)\sin^4{\theta_W}}\nonumber \\
& - &\frac{8 \beta ^2 E_e^4 e^4  \left(\left(3 \beta ^4-2 \beta ^2+3\right) \eta_T 
	\xi_T  (a_f^2-v_f^2)  \cos (2 \phi -\phi_ {e^{-}}+\phi_ {e^{+}})\right)}{3 M^4_W \left(s-M^2_Z\right)^2\sin^4{\theta_W}}
\Bigg).
\end{eqnarray}
After putting the values of masses, couplings and energy of the beam in the non-oscillatory part of the above equation, while keeping the coefficient of the oscillatory part as a function of $\eta_T$ , $\xi_T$ , and $f(\beta_W)$, we find the following expression for the differential cross section
\begin{eqnarray}\label{eq:phi-sm}
\frac{d\sigma}{d\phi} = 427.968 + \eta_T  \xi_T f(\beta)\cos (2 \phi -\phi_ {e^-}+\phi_ {e^+}),
\end{eqnarray}
where
\begin{eqnarray}
f(\beta) & = & \frac{\left(0.0261478 \beta ^8-0.104591 \beta ^6+0.156887 \beta ^4-0.104591 \beta ^2+0.0261478\right) \log \left((1-\beta )^2\right)}{M^4_W} \nonumber \\
& - & \frac{\left(0.0522956 \beta ^8-0.209182 \beta ^6+0.313774 \beta ^4-0.209182 \beta ^2+0.0522956\right) \log (1+\beta )}{M^4_W} \nonumber \\
& - & \frac{0.105566 \beta ^7-0.174968 \beta ^5+0.175293 \beta ^3-0.104591 \beta }{M^4_W}.
\end{eqnarray}

\subsubsection{\rm\bf IDM: $ e^-(p_1,s_1) e^+(p_2,s_2) \to H^-(k_2) H^+(k_1) $} 
The amplitudes, in this case, for the $s$-channel diagrams containing photon and $Z$ boson are given below as
\begin{equation}
\mathcal{M}_{\gamma}(s_1,s_2) = - \frac{i e^2 }{s} \,  \bar{u}(s_2,p_2)  \gamma^{\mu} u(s_1,p_1) (k_1 - k_2)_\mu,
\end{equation}
\begin{equation}
\mathcal{M}_{Z}(s_1,s_2) = \frac{i g_Z (g_2 \cos{\theta_W} -g_1 \sin{\theta_W} )   }{4\, (s-M^2_Z)} \,  \bar{u}(s_2,p_2)  \gamma^{\mu} (v_f - a_f \gamma_5) u(s_1,p_1) (k_1 - k_2)_\mu.
\end{equation}

The square amplitude for the process is given by
\begin{equation}
| \mathcal{M}|^2 = \sum_{s_i}\mathcal{M} (s_1,s_2) \mathcal{P}_{e^{-}}(s_1,s_3) \mathcal{P}_{e^{+}}(s_2,s_4) \mathcal{M}^{\dagger} (s_3,s_4),  
\end{equation}
where
\begin{equation}
\mathcal{M}  = \mathcal{M}_{\gamma} + \mathcal{M}_{Z}. 
\end{equation}
The expressions for square amplitudes of the two diagrams along with their interference are written as 
\begin{equation}
| \mathcal{M}|^2 = |\mathcal{M}_\gamma|^2  +|\mathcal{M}_Z|^2 
+2Re\left(\mathcal{M}_\gamma\mathcal{M}_Z^\dagger\right) .
\end{equation}
Here, 
\begin{eqnarray}\label{eq:idm-photon}
|\mathcal{M}_\gamma|^2 &=& \frac{1}{4} \beta ^2 s\theta ^2 e^4  e^{-i (\phi_{e^{-}} +\phi_{e^{+}} )} \Bigg(2 e^{i (\phi_{e^{-}}
	+\phi_{e^{+}} )}-i \eta_T  \xi_T  \sin (2 \phi ) \left(e^{2 i
	\phi_{e^{-}} }-e^{2 i \phi_{e^{+}} }\right) \nonumber \\
&+&\eta_T  \xi_T  \cos (2 \phi ) \left(e^{2 i \phi_{e^{-}} }+e^{2 i \phi_{e^{+}} }\right)\Bigg),
\end{eqnarray}
\begin{eqnarray}
|\mathcal{M}_Z|^2 & = & \frac{1}{\left(s-M^2_Z\right)^2\tan^2{2\theta_W}} \Bigg[\beta ^2 s\theta ^2 E_e^4 g_Z^2 e^{-i (\phi_{e^{-}} +\phi_{e^{+}} )} e^2  \left(2 
\left(a_f ^2+v_f ^2\right) e^{i (\phi_{e^{-}} +\phi_{e^{+}} )}\right)\nonumber \\
&  - &\beta ^2 \left(s\theta ^2\right) E_e^4 g_Z^2 e^{-i (\phi_{e^{-}} +\phi_{e^{+}} )} e^2\left(\eta_T  \xi_T  \left(a_f ^2-v_f ^2\right) \cos (2 \phi ) \left(e^{2 i \phi_{e^{-}} }+e^{2 i \phi_{e^{+}}
}\right)\right) \nonumber \\ 
&+&i\beta ^2 s\theta ^2 E_e^4 g_Z^2 e^{-i (\phi_{e^{-}} +\phi_{e^{+}} )} e^2  \left(\eta_T  \xi_T  \left(a_f ^2-v_f ^2\right) \sin (2 \phi ) \left(e^{2 i \phi_{e^{-}}
}-e^{2 i \phi_{e^{+}} }\right)\right)\Bigg], 
\end{eqnarray}
\begin{eqnarray}
Re\left(\mathcal{M}_\gamma\mathcal{M}_Z^\dagger\right)& = & \dfrac{1}{2\left(s-M^2_Z\right)\tan{2\theta_W}}\Bigg[\beta ^2 \left(c\theta ^2-1\right) E_e^2 e^3 g_Z  v_f e^{-i (\phi_{e^{-}} +\phi_{e^{+}} )} 
\left(2 e^{i (\phi_{e^{-}} +\phi_{e^{+}} )}\right) \nonumber \\
&  +& \beta ^2 \left(s\theta ^2\right) E_e^2 e^3 g_Z  v_f e^{-i (\phi_{e^{-}} +\phi_{e^{+}} )} 
\left(i \eta_T  \xi_T  \sin (2 \phi ) \left(e^{2 i \phi_{e^{-}} }-e^{2 i \phi_{e^{+}} }\right)\right) \nonumber \\		 
&+& \beta ^2 \left(s\theta ^2\right) E_e^2 e^3 g_Z  v_f e^{-i (\phi_{e^{-}} +\phi_{e^{+}} )} 
\left(\eta_T  \xi_T  \cos (2 \phi ) \left(e^{2 i \phi_{e^{-}}
}+e^{2 i \phi_{e^{+}} }\right)\right) \Bigg]. 
\end{eqnarray}
In this case, the $\beta$ is the boost for the $H^\pm$.
From the above square amplitude, we obtain the differential cross section, after integrating over the variable $\theta$, as
\begin{eqnarray}
\frac{d\sigma}{d\phi}& = & \frac{  3.894 \times 10^{11}}{64 \pi^2 s}\Bigg(
\frac{8 \beta ^3 E_e ^4 g_Z^2 e^2 \left(a_f ^2-\eta_T  \xi_T  (a_f -v_f ) (a_f +v_f ) \cos (2 \phi
	-\phi_{e^{-}}+\phi_{e^{+}})+v_f ^2\right)}{3 \left(s-M^2_Z\right)^2\tan^2{2\theta_W}} \nonumber \\
&+&	\frac{2}{3} \beta ^3 e^4  (\eta_T  \xi_T  \cos (2 \phi -\phi_{e^{-}}+\phi_{e^{+}})+1)-\frac{8 \beta ^3 E_e ^2 e^3 g_Z 
	v_f   (\eta_T  \xi_T  \cos (2 \phi -\phi_{e^{-}}+\phi_{e^{+}})+1)}{3 \left(s-M^2_Z\right)\tan{2\theta_W}}\Bigg).\nonumber\\
\end{eqnarray}
With the values of masses, couplings and energy of the beam, we find the differential cross section to be a function of  $\eta_T$ , $\xi_T$ , and $f(\beta)$ as follows
\begin{eqnarray}\label{eq:phi-idm}
\frac{d\sigma}{d\phi} =4.64277 \beta ^3 +  3.42967\,  \eta_T  \xi_T \beta ^3 \cos (2 \phi -\phi_{e^{-}} +\phi_{e^{+}} ).
\end{eqnarray}

\subsubsection{\rm\bf MSSM: $ e^-(p_1,s_1) e^+(p_2,s_2) \to \chi^-(k_1,s_3) \chi^+(k_2,s_4)$} 
The amplitudes of two $s$-channel processes have the following expressions
\begin{equation}
\mathcal{M}_{\gamma}(s_1,s_2,\lambda_3,\lambda_4) =  \frac{i e^2 }{s} \,  \bar{v}(s_2,p_2)  \gamma_{\mu} u(s_1,p_1) \bar{u}_\chi(\lambda_3,k_1)  \gamma^{\mu} v_\chi(\lambda_4,k_2),
\end{equation}
\begin{equation}
\mathcal{M}_{Z}(s_1,s_2,\lambda_3,\lambda_4) = \frac{i g^2_Z }{4\,(s-M^2_Z)} \,  \overline{v}(s_2,p_2)  \gamma^{\mu} (v_f - a_f \gamma_5) u(s_1,p_1) \bar{u}_\chi(\lambda_3,k_1)  \gamma_{\mu} (v_\chi - a_\chi \gamma_5) v_\chi(\lambda_4,k_2).
\end{equation}

The square of the amplitude is written as
\begin{equation}
| \mathcal{M}|^2 = \sum_{s_i,\lambda_i}\mathcal{M} (s_1,s_2,\lambda_3,\lambda_4) \mathcal{P}_{e^{-}} (s_1,s_3)\mathcal{P}_{e^{+}}(s_2,s_4) \mathcal{M}^{\dagger} (s_3,s_4,\lambda_3,\lambda_4), 
\end{equation}
where 	
\begin{equation}
\mathcal{M}  = \mathcal{M}_{\gamma} + \mathcal{M}_{Z}. 
\end{equation}
The expressions for the square amplitude of two $s$-channel diagrams and their cross term are given  as
\begin{equation}
| \mathcal{M}|^2 = |\mathcal{M}_\gamma|^2  +|\mathcal{M}_Z|^2 
+2Re\left(\mathcal{M}_\gamma\mathcal{M}_Z^\dagger\right) .
\end{equation}
Here, 
\begin{eqnarray}\label{eq:mssm-photon}
|\mathcal{M}_\gamma|^2 & = & \frac{e^4}{2}  \left( e^{-i (2 \phi +\phi_{e^{-}} +\phi_{e^{+}} )} \beta ^2 \left(c\theta ^2-1\right) \eta_T 
\xi_T \left( e^{2 i (2 \phi +\phi_{e^{+}} )}+ e^{2 i \phi_{e^{-}} }\right)  	-     \left(2 \left(\beta ^2 s\theta
^2-2\right) \right)\right),  \nonumber\\ 
\end{eqnarray}

\begin{eqnarray}
|\mathcal{M}_Z|^2 & = &\frac{1}{{\left(s-M^2_Z\right)^2}} \Bigg[E_e^4 g_z^2 v_f^2 \left(a_\chi^2 \beta ^2 \left(c\theta ^2+1\right)+v_\chi ^2 \left(\beta ^2 \left(c\theta
^2-1\right)+2\right)\right) \nonumber \\
& +&E_e^4 g_z^2 \left(2 a_f^2 \left(a_\chi^2 \beta ^2 \left(c\theta ^2+1\right)+v_\chi ^2 \left(\beta ^2 \left(c\theta
^2-1\right)+2\right)\right)+16 a_f a_\chi \beta  c\theta  v_f v_\chi \right) \nonumber \\
&   + &\left(a_f^2-v_f ^2\right) E_e^4 g_z^2 e^{-i (2 \phi +\phi_{e^{-}}+\phi_{e^{+}})}\beta ^2 \left(s\theta ^2\right) \eta_T  \xi_T 
\left(a_\chi^2+v_\chi ^2\right) \left( e^{2 i (2 \phi +\phi_{e^{+}})}+ e^{2 i \phi_{e^{-}}}
\right) 
\Bigg],   \nonumber\\
\end{eqnarray}
\begin{eqnarray}
Re\left(\mathcal{M}_\gamma\mathcal{M}_Z^\dagger\right) & = & \frac{1}{2\left(s-M^2_Z\right)}\Bigg[ 2  E_e^2 e ^2 g_z  \beta ^2 \left(c\theta
^2-1\right) \eta_T  \xi_T 
v_f v_\chi e^{2 i \phi_{e^{-}}} \nonumber \\
&+& 2  E_e^2 e ^2 g_z  \left(2 a_f  a_\chi \beta  c\theta +v_f v_\chi
\left(\beta ^2 \left(c\theta ^2-1\right)+2\right)\right) \nonumber\\		
& +& E_e^2 e ^2 g_z  e^{-i (2 \phi +\phi_{e^{-}}+\phi_{e^{+}})} \left(\beta ^2 \left(c\theta
^2-1\right) \eta_T  \xi_T  v_f v_\chi e^{2 i (2 \phi +\phi_{e^{+}})}\right)\Bigg].
\end{eqnarray}
Here, the $\beta$ stands for the boost of $\chi^\pm$.
Using the above matrix amplitude square the differential cross section  is found to be
\begin{eqnarray}
\frac{d\sigma}{d\phi}& = &\frac{ 3.894 \times 10^{11} \, \beta}{64\, \pi^2 s}\Bigg(\frac{8 E_e ^2 e^2 g_z  v_f  v_\chi  \left(\beta ^2 \eta_T  \xi_T  \cos (2 \phi -\phi_{e^{-}}+\phi_{e^{+}})+\beta
	^2-3\right)}{3  \left(s - M^2_Z\right)} \nonumber \\
& +&  \frac{4  E_e ^4 g^2_z \left(\beta ^2 \eta_T  \xi_T  (a^2_f -v^2_f ) (a_f +v_f ) \left(a_\chi ^2+v_\chi ^2\right) \cos (2 \phi -\phi_{e^{-}}
	+\phi_{e^{+}})\right)}{3  \left(s - M^2_Z\right)^2} \nonumber \\			
& - &\frac{4}{3} e^4  \left(\beta ^2 \eta_T  \xi_T  \cos (2 \phi -\phi_{e^{-}}+\phi_{e^{+}})+\beta
^2-3\right) \nonumber \\
&+&  \frac{4  E_e ^4 g^2_z \left(\left(a_f ^2+v_f ^2\right) \left(\left(3-\beta ^2\right) v_\chi ^2+2 a_\chi ^2
	\beta ^2\right)\right)}{3  \left(s - M^2_Z\right)^2} \Bigg).
\end{eqnarray}

Substituting the values of  all the couplings, masses and beam energy except the $m_\chi$, we obtain the  expression for the differential cross section as 
\begin{eqnarray}\label{eq:phi-mssm}
\frac{d\sigma}{d\phi}  = 52.4371 \beta-17.457 \beta ^3 -0.966784 \beta ^3 \eta _T \xi _T \cos \left(2 \phi + \phi _{e^+}-\phi _{e^-} \right).
\end{eqnarray}

\providecommand{\href}[2]{#2}\begingroup\raggedright\endgroup	
	
\end{document}